%% file: sefv.tex
\documentclass[aps,prd,superscriptaddress,floatfix,preprint,11pt]{revtex4-1}
\makeatletter
\def\l@subsubsection#1#2{}
\makeatother
\usepackage[english]{babel}
\usepackage[utf8]{inputenc}
\usepackage[T1]{fontenc}
\usepackage{amsmath,amssymb,braket,slashed,mathrsfs}
\usepackage{pifont}

\usepackage{bm}

\usepackage[squaren,Gray,cdot,binary]{SIunits}
\usepackage{graphicx}
\graphicspath{{./figures/}}
\usepackage{color,hyperref}
\usepackage{minibox}
\usepackage{cleveref}
\usepackage{marginnote}

\crefformat{section}{Sec.~#2#1#3}
\crefmultiformat{section}{Sec.~#2#1#3}
    { and~#2#1#3}{, #2#1#3}{ and~#2#1#3}
\crefrangeformat{section}{Sec.~#3#1#4\,--\,#5#2#6}
\crefformat{subsection}{Sec.~#2#1#3}
\crefmultiformat{subsection}{Sec.~#2#1#3}
    { and~#2#1#3}{, #2#1#3}{ and~#2#1#3}
\crefrangeformat{subsection}{Sec.~#3#1#4\,--\,#5#2#6}
\crefformat{subsubsection}{Sec.~#2#1#3}
\crefmultiformat{subsubsection}{Sec.~#2#1#3}
    { and~#2#1#3}{, #2#1#3}{ and~#2#1#3}
\crefrangeformat{subsubsection}{Sec.~#3#1#4\,--\,#5#2#6}
\crefformat{figure}{Fig.~#2#1#3}
\crefmultiformat{figure}{Figs.~#2#1#3}
    { and~#2#1#3}{, #2#1#3}{ and~#2#1#3}
\crefrangeformat{figure}{Figs.~#3#1#4\,--\,#5#2#6}
\crefformat{table}{Table~#2#1#3}
\crefmultiformat{table}{Tables~#2#1#3}
    { and~#2#1#3}{, #2#1#3}{ and~#2#1#3}
\crefrangeformat{table}{Tables~#3#1#4\,--\,#5#2#6}
\crefformat{equation}{Eq.~(#2#1#3)}
\crefmultiformat{equation}{Eqs.~(#2#1#3)}{ and~(#2#1#3)}{, (#2#1#3)}{ and~(#2#1#3)}
\crefrangeformat{equation}{Eqs.~(#3#1#4)--(#5#2#6)}
                
\definecolor{linkcolor}{rgb}{.17578125,.1875,.5703125}
\hypersetup{
    pdfstartview={FitH},                         
    pdftitle={Volume dependence of the electromagnetic 
              self-energy function},                            
    pdfauthor={Antonin Portelli},                         
    pdfsubject={Physics article},                              
    colorlinks=true,                             
    linkcolor=linkcolor,                         
    citecolor=linkcolor,                         
    filecolor=black,                             
    urlcolor=linkcolor      						 
}

\newcommand{\eg}{\textit{e.g.}~}
\newcommand{\ie}{\textit{i.e.}~}

\DeclareMathOperator{\arcsinh}{arcsinh}
\DeclareMathOperator{\arctanh}{arctanh}
\renewcommand{\bar}{\overline}
\DeclareMathOperator{\bigo}{\mathcal{O}}

\newcommand{\diff}{\mathrm{d}}
\newcommand{\Diff}{\mathcal{D}}
\renewcommand{\epsilon}{\varepsilon}
\newcommand{\F}{\mathcal{F}}

\newcommand{\QED}{\mathrm{QED}}
\newcommand{\R}{\mathbb{R}}
\renewcommand{\Re}{\mathrm{Re}}
\DeclareMathOperator{\Res}{Res}
\DeclareMathOperator{\SU}{SU}
\DeclareMathOperator{\U}{U}
\DeclareMathOperator*{\sump}{\sideset{}{^\prime}\sum}
\newcommand{\T}{\mathbb{T}}
\renewcommand{\tilde}{\widetilde}
\DeclareMathOperator{\sign}{sign}

\newcommand{\Z}{\mathbb{Z}}

\newcommand{\df}{\delta}
\newcommand{\db}{\delta^*}
\newcommand{\cdf}{\nabla}
\newcommand{\cdb}{\nabla^*}
\newcommand{\Dt}{\delta^2}
\newcommand{\cDt}{\nabla^2}
\newcommand{\suml}[1]{\sum_{#1\in\Lambda^4}}

\newcommand{\nsp}{\mathbf{n}}
\newcommand{\usp}{\hat{\mathbf{n}}}
\newcommand{\ksp}{\mathbf{k}}
\newcommand{\psp}{\mathbf{p}}
\newcommand{\vel}{\mathbf{v}}
\newcommand{\om}{\omega(\psp)}
\newcommand{\omk}{\omega(\psp-\ksp)}

\newcommand{\vdnu}{\vel\cdot\usp}
\newcommand{\vdku}{\vel\cdot\hat{\ksp}}
\newcommand{\vudnu}{\hat{\vel}\cdot\usp}
\newcommand{\gv}{\gamma(|\vel|)}
\newcommand{\qedl}{\mathrm{QED}_{\mathrm{L}}}
\newcommand{\qedtl}{\mathrm{QED}_{\mathrm{TL}}}
\newcommand{\qedc}{\mathrm{QED}_{\mathrm{C}}}
\newcommand{\qedm}{\mathrm{QED}_{\mathrm{M}}}

\begin{document}
  \title{
  Theoretical aspects of quantum electrodynamics\\ 
  in a finite volume with periodic boundary conditions
 }
  
  \author{Z. Davoudi}
  \affiliation{Department of Physics, 
  University of Maryland, College Park, MD 20742, USA}
  \affiliation{RIKEN Center for Accelerator-based Sciences, Wako 351-0198, Japan}
  \author{J. Harrison}
  \affiliation{School of Physics and Astronomy, 
  University of Southampton, Southampton SO17 1BJ, United Kingdom}
  \author{A. Jüttner}
  \affiliation{School of Physics and Astronomy, 
  University of Southampton, Southampton SO17 1BJ, United Kingdom}
  \author{A. Portelli}
  \email[corresponding author, ]{antonin.portelli@ed.ac.uk}
  \affiliation{School of Physics and Astronomy, 
  University of Edinburgh, Edinburgh EH9 3FD, United Kingdom}
  \author{M.J. Savage}
  \affiliation{Institute for Nuclear Theory, 
  University of Washington, Seattle, WA 98195, USA}
  \preprint{INT-PUB-18-051}
  \preprint{UMD-PP-018-07}
  \begin{abstract}
    First-principles studies of strongly-interacting hadronic systems using
    lattice quantum chromodynamics (QCD) have been complemented in recent years
    with the inclusion of quantum electrodynamics (QED). The aim is to confront
    experimental results with more precise theoretical determinations, \eg for
    the anomalous magnetic moment of the muon and the CP-violating parameters in
    the decay of mesons. Quantifying the effects arising from
    enclosing QED in a finite volume remains a primary target of investigations.
    To this end, finite-volume corrections to hadron masses in the presence of
    QED have been carefully studied in recent years. This paper extends such
    studies to the self-energy of moving charged hadrons, both on and away from
    their mass shell. In particular, we present analytical results for leading
    finite-volume corrections to the self-energy of spin-0 and
    spin-$\frac{1}{2}$ particles in the presence of QED on a periodic hypercubic
    lattice, once the spatial zero mode of the photon is removed, a framework
    that is called $\qedl$. By altering modes beyond the zero mode, an
    improvement scheme is introduced to eliminate the leading finite-volume
    corrections to masses, with potential applications to other hadronic
    quantities. Our analytical results are verified by a dedicated numerical
    study of a lattice scalar field theory coupled to $\qedl$. Further, this
    paper offers new perspectives on the subtleties involved in applying
    low-energy effective field theories in the presence of $\qedl$, a
    theory that is rendered non-local with the exclusion of the spatial zero
    mode of the photon,  clarifying recent discussions on this matter.  
  \end{abstract}
  \maketitle
  \tableofcontents
  \section{Introduction}
  \label{sec:introduction}
  \input{sections/introduction}
  \section{$\qedl$ in the path integral formalism}
  \label{sec:qedl}
  \input{sections/qedl}
  \section{Finite-size effects in $\qedl$: the self-energy function}
  \label{sec:fv}
  \input{sections/fv}
  \section{Infrared improvement of the $\qedl$ theory}
  \label{sec:irimp}
  \input{sections/improvement}
  \section{Numerical study through simulations of lattice scalar QED}
  \label{sec:lattice}
  \input{sections/lattice}
  \section{Low-Energy Effective Field Theories}
  \label{sec:EFT}
  \input{sections/eft}
  \section{Summary, Conclusions and Outlook}
  \label{sec:sco}
  \input{sections/sco}
  \begin{acknowledgments}
    The authors would like to especially thank Peter Boyle for useful
    conversations and a critical read of the manuscript. A.P. would like to
    thank the Institute for Nuclear Theory (INT) of the University of Washington
    (UW) for its very warm welcome. A.P. would also like to thank Chris
    Sachrajda for useful discussions. M.J.S. would like to thank Silas Beane and
    Brian Tiburzi for helpful discussions. Numerous concepts presented here
    emerged from discussions during A.P.'s extended stay at the INT. Numerical
    lattice QED computations presented in this work were performed on the Hyak
    High performance Computing and Data Ecosystem at the UW, the IRIDIS High
    Performance Computing Facility at the University of Southampton, and on
    DiRAC equipment, including the Extreme Scaling service Tesseract in
    Edinburgh. DiRAC is part of the UK National E-Infrastructure. The simulation
    software used in this project was developed as part of the Grid \& Hadrons
    libraries (\url{https://github.com/paboyle/Grid}), which are free software
    distributed under the General Public License version 2. A.P. is supported in part
    by UK STFC grants ST/L000458/1 and ST/P000630/1, and the European Research
    Council (ERC) under the European Union's Horizon 2020 research and
    innovation program under grant agreement No 757646. Z.D. was partly
    supported by the Maryland Center for Fundamental Physics. A.J. received
    funding from STFC consolidated grant ST/P000711/1 and from the European
    Research Council under the European Union's Seventh Framework Program
    (FP7/2007- 2013) / ERC Grant agreement 279757. M.J.S. is supported at the
    INT by US Department of Energy grant number DE-FG02-00ER41132. J.H. is
    supported by the EPSRC Centre for Doctoral Training in Next Generation
    Computational Modelling grant EP/L015382/1.
  \end{acknowledgments}
  \appendix
  \section{Derivation of the general finite-volume formula}
  \label{sec:masterfv}
  \input{sections/masterfv}
  \section{Numerical computation of the finite-volume coefficients}
  \label{sec:cj}
  \input{sections/cj}
  \section{Velocity suppression of the harmonic coefficients $a_{klm}(\vel)$}
  \label{sec:aklm}
  \input{sections/aklm}
  \section{Time-momentum representation of lattice scalar correlators}
  \label{sec:latticetmom}
  \input{sections/latticetmom}
  \section{Parameters used for fits of the lattice on-shell scalar self-energy}
  \label{sec:emfit}
  \input{sections/emfit}
  \pagebreak
  \bibliography{sefv}
\end{document}

%% file: sections/introduction.tex
State-of-the-art simulations of QCD reliably predict a number of spectral
quantities and hadronic matrix elements with a precision below the percent
level, see for instance the review by the Flavour Lattice Averaging Group
(FLAG)~\citep{Aoki:2016frl}. Most of the results listed by FLAG have been
obtained within an isospin-symmetric QCD, \ie, with equal light quark masses
and ignoring electromagnetic interactions. A logical next step in the continuous
improvement of calculations of spectra, matrix elements and scattering
amplitudes is the inclusion of isospin breaking effects, which by naive power
counting are expected to contribute at the percent level and are hence becoming
significant. First efforts in this direction date back over two
decades~\citep{Duncan:1996xy}, and interest in this field has picked up
considerably over the last few years. First results are now available, in
particular, for spectral quantities
~\citep{Blum:2007cy,Blum:2010ym,Ishikawa:2012ix,Aoki:2012st,
Borsanyi:2013lga,Borsanyi:2014jba,Portelli:2015wna,Horsley:2015eaa,
Horsley:2015vla,Fodor:2016bgu,deDivitiis:2013xla,Giusti:2017dmp,
Basak:2018yzz,Hansen:2018zre}, and progress is being made in matrix elements and
scattering and decay
amplitudes~\citep{Blum:2017cer,Blum:2018mom,Carrasco:2015xwa,Lubicz:2016xro,
Giusti:2017dwk,Christ:2017pze,Giusti:2017jof,Boyle:2017gzv}.

A fundamental difficulty with the formulation of QED concerns Gauss' law, which
implies that gauge-invariant charged states cannot exist in a finite volume with
periodic boundary conditions. As will be discussed later, this problem is
related to the occurrence of global photon zero modes. Various proposals exist
on how to deal with the zero-mode problem: In
$\qedtl$~\citep{Duncan:1996xy,Duncan:1996be,Borsanyi:2013lga,Ishikawa:2012ix,
Aoki:2012st,deDivitiis:2013xla,Fodor:2016bgu} the global photon zero mode is
removed from the dynamics, while in
$\qedl$~\citep{Hayakawa:2008an,Davoudi:2014qua,Fodor:2015pna,Blum:2007cy,
Blum:2010ym,Ishikawa:2012ix,Giusti:2017dwk,Giusti:2017dmp,Giusti:2017jof,
Blum:2018mom,Boyle:2017gzv,Lee:2015rua,Matzelle:2017qsw} the photon zero mode is
removed individually on every time slice. Locality is violated in both cases.
$\qedl$ does, however, allow for a transfer matrix to be constructed that is
reflection-positive, making it the preferred choice. Understanding and
controlling the implications of the locality violation remains an important
task. Alternatives to subtracting the zero mode have also been suggested,
allowing the locality to be preserved in any finite volume. In massive QED,
called $\qedm$, a small photon mass is introduced as an IR
regulator~\citep{Endres:2015gda} and physical QED results are extracted from an
extrapolation to the zero photon mass~\citep{Endres:2015gda, Bussone:2017xkb}.
Charge conjugation boundary
conditions~\citep{Polley:1993bn,Wiese:1991ku,
Kronfeld:1992ae,Kronfeld:1990qu,Lucini:2015hfa,Hansen:2018zre} have been
proposed as a way to allow the construction of gauge-invariant charged states in
a finite volume. In this construction, called $\qedc$, charge and flavor
conservation are partially broken by the boundary conditions and these effects
need to be controlled at any finite volume. $\qedm$ and $\qedc$ provide
promising avenues towards local simulations of QCD+QED.

All approaches introduced above suffer from large finite-volume effects induced by
the absence of a mass gap (or the small photon mass in the case of $\qedm$).
Understanding these effects analytically has been the subject of a series of
articles~\citep{Borsanyi:2014jba,Davoudi:2014qua,Lubicz:2016xro,Lucini:2015hfa,Endres:2015gda}.
These finite-volume effects are power suppressed in $L$ for massless photons,
where $L$ denotes the extent of the finite cubic volume, and their precise form
depends on which formulation of QED in a finite volume is implemented. In a
number of cases, the leading coefficients of the power expansion are universal.
In such cases, the large-distance limit of the finite-volume effects are
equivalent to those of point particles and can be computed and corrected for
analytically, for instance by means of perturbative calculations in
scalar/fermionic QED or in effective field theories, see e.g,
Refs.~\citep{Borsanyi:2014jba,Davoudi:2014qua}.

The main objective of this paper is to provide a simple and versatile recipe for
computing universal finite-volume effects analytically. Some general remarks on
$\qedtl$ and $\qedl$ are provided, and it is shown how each of these theories
are quantized in the path-integral formalism. This is followed by the core part
of this paper, namely a proposal for a systematic computation of QED
finite-volume effects in terms of a large-volume expansion. Existing results for
the finite volume effects on spectral quantities are reproduced and are further
extended to on-shell and off-shell finite-volume effects in moving frames, where
new rotational symmetry breaking effects are observed and quantified. This
recipe is applied to the self-energy of charged fundamental particles with spins
0 and $\frac12$. This paper will be followed by another
work~\citep{Bijnens:2018} by some of the authors, applying the same procedure to
compute the finite-volume effects on the electromagnetic corrections to the
hadronic vacuum polarization. Inspired by Symanzik's improvement program aimed
at reducing lattice-cutoff effects, a proposal is made for improving the
infrared behavior of the $\qedl$ theory, and is shown to remove universal
finite-volume effects by modifying individual momentum modes in the photon
action. All the analytical predictions for scalar QED are confirmed by
high-statistics simulations. Features of these simulations, such as excited
states contributions and the signal-to-noise degradation in boosted systems, are
discussed. We conclude this paper with clarifying remarks on effective theories
of $\qedl$. This discussion provides insights into the origin of the discrepancy
between full $\qedl$ and its corresponding effective theories for the
higher-order finite-volume QED effects. New local operators with
volume-dependent coefficients are introduced into the effective theories without
the need to include the anti-particle modes.

%% file: sections/qedl.tex
In this section we retrace the construction of $\qedl$, first introduced in
Ref.~\citep{Hayakawa:2008an} from the point of view of path
integral quantization. While most readers will be familiar with the definition
of $\qedl$, we provide a more formal definition in terms of the path integral as
the starting point for perturbative expansions or lattice discretizations
discussed in later sections. We start by making general observations about QED
in a finite volume and then introduce $\qedtl$ and $\qedl$.
\subsection{Periodic fields and zero-mode singularities}
The infinite-volume Euclidean Maxwell action in Feynman gauge is
\begin{align}
  S[A_{\mu}]=&\int\diff^4 x\left\{
    \frac14F_{\mu\nu}(x)F_{\mu\nu}(x)
    +\frac12[\partial_{\mu}A_{\mu}(x)]^2
  \right\}\notag
  \\
  =&-\frac12\int\diff^4 x\,A_{\mu}(x)\,\partial^2
  A_{\mu}(x)\,,
\end{align}
where $F_{\mu\nu}=\partial_{\mu}A_{\nu}-\partial_{\nu}A_{\mu}$ is the field
strength tensor and $A_{\mu}$ is the $\U(1)$ gauge potential. In momentum space,
this action takes the convenient form
\begin{equation}
   S[\hat{A}_{\mu}]=\frac12\int\frac{\diff^4 k}{(2\pi)^4}
  \ k^2\  {\textstyle\sum_{\mu}}|\hat{A}_{\mu}(k)|^2\,,
  \label{eq:momspaceinfvact}
\end{equation}
where the following Fourier transform normalization is used
\begin{equation}
  \hat{A}_{\mu}(k)=\int\diff^4 x\,A_{\mu}(x)e^{-ik\cdotp x}\,.
\end{equation}
The theory is quantized by means of the Euclidean path integral. The vacuum
expectation value of operator $O$ in the absence of matter fields is defined as
\begin{equation}\label{eq:pint}
  \braket{O}=\frac{1}{\mathcal{Z}}\int\Diff A_{\mu}\,
  O[A_{\mu}]\exp(-S[A_{\mu}])\,,
\end{equation}
where $\mathcal{Z}=\int\Diff A_{\mu}\exp(-S[A_{\mu}])$ is the partition
function. In this free theory, any expectation value can be expressed in terms
of the photon propagator
\begin{equation}\label{eq:Dmunu_inf}
  D_{\mu\nu}^{(\infty)}(x-y)=-\delta_{\mu\nu}(\partial^2)^{-1}\delta(x-y)=
  \int\frac{\diff^4k}{(2\pi)^4}\frac{\delta_{\mu\nu}}{k^2}e^{ik\cdotp(x-y)}\,.
\end{equation}
The inverse Laplacian is defined unambiguously in infinite volume
since its zero mode constitutes a set of measure zero within the continuous
spectrum of the operator.

Now consider the above path integral in a finite volume of spacetime with
spatial dimensions of equal length $L$ and a time extent $T$. Here, periodic
boundary conditions are imposed, and the physical space-time volume is denoted
by $\T^4$. For the sake of simplicity and since we are only interested in
long-distance effects, spacetime is assumed to be continuous. Momentum is
quantized  on $\T^4$, and the Fourier transform is defined by
\begin{equation}
  \hat{f}(k)=\int_{\T^4}\diff^4x\, f(x)e^{-ik\cdotp x}
  \qquad\text{and}\qquad f(x)=\frac{1}{TL^3}\sum_{k\in\hat{\T}^4}
  \hat{f}(k)e^{ik\cdotp x}\,,
\end{equation}
where $\hat{\T}^4$ is the discrete set of vectors of the form
$(\frac{2\pi}{T}n_0,\frac{2\pi}{L}\nsp)$ where $n=(n_0,\nsp)$ is a four-vector
with integer components. On $\T^4$, one could attempt to define QED in terms of
the continuum limit of a discretized version of the momentum-space action given
in~\cref{eq:momspaceinfvact}:
\begin{equation}
  S[\hat{A}_{\mu}]=\frac1{2TL^3}\sum_{k\in\hat{\T}^4}
  k^2\,{\textstyle\sum_{\mu}}|\hat{A}_{\mu}(k)|^2\,.
  \label{eq:naivefvact}
\end{equation}
In this case,  the zero mode of the Laplacian is a significant, isolated mode,
and the finite-volume equivalent of~\cref{eq:Dmunu_inf},
\begin{equation}
  D_{\mu\nu}(x-y)=
  \frac1{TL^3}\sum_{k\in\hat{\T}^4}
  \frac{\delta_{\mu\nu}}{k^2}e^{ik\cdotp(x-y)}\,,
\end{equation}
is ill-defined because of the singular $k=0$ term in the sum. In other words,
the Laplacian is not invertible. Now let us define a \emph{shift transformation}
through
\begin{equation}
  A_{\mu}(x)\mapsto A_{\mu}^{b}(x)=A_{\mu}(x)+\frac{b_{\mu}}{TL^3}\,,
  \label{eq:shifttr}
\end{equation}
where $b_{\mu}$ is a constant four-vector with a mass dimension equal to $-3$.
In momentum space, this shift transformation becomes
\begin{equation}
  \hat{A}_{\mu}^{b}(k)=\hat{A}_{\mu}(k) + b_{\mu} \delta_{k,0}\,,
\end{equation}
\ie it modifies the zero-mode of the EM potential. The action in
~\cref{eq:naivefvact} is invariant under such shift transformations and hence
the Laplacian is not invertible. One can in fact show that for periodic boundary
conditions shift transformations span the whole nullspace of the Laplacian. 

One can observe that in~\cref{eq:shifttr}, the shift $b_{\mu}$ can in principle
be written as the derivative of a linear function $\omega$, which makes the
shift transformation a gauge transformation. To be smooth, the associated
$\U(1)$ transformation $\exp(i\omega)$ requires $b_{\mu}$ to fullfil some
trivial quantization condition. Such function is not homotopic to the identity
on $\U(1)$, and is part of the class of ``large'' gauge transformations. As it
will be discussed at length in the next sections, the redundancy associated to
these transformations cannot be fixed using a local gauge fixing prescription.
\subsection{The $\qedtl$ theory}
The shift symmetry described in the previous section is somewhat similar to the
problem in gauge theory that motivates gauge fixing: the action of the theory
has an internal symmetry which generates a singular redundancy in the space of
field configurations. In the case of shift symmetry, this redundancy can be
eliminated by using the same Fadeev and Popov procedure~\citep{Faddeev:1967fc}
that is used to implement gauge fixing in the path integral formalism of gauge
theories. We start by inserting
\begin{equation}
  1=\int\diff b_{\mu}\
  \delta[{\textstyle\int_{\T^4}}\diff^4x\,A_{\mu}^b(x)]\,,
\end{equation}
into the path integral in~\cref{eq:pint},
\begin{equation}
    \braket{O}=\frac{1}{\mathcal{Z}}\int\Diff A_{\mu}
    \int\diff b_{\mu}\,
    \delta\left[{\textstyle\int_{\T^4}}\diff^4x\,A_{\mu}^b(x)\right]
    O[A_{\mu}]\exp(-S[A_{\mu}])\,.
\end{equation}
Using the invariance of the action under $A_{\mu}\mapsto A_{\mu}^b$ and assuming
the same property for the operator, the  infinite factor of $\int\diff b_{\mu}$
can be canceled between the numerator and the denominator to obtain
\begin{equation}
    \braket{O}=\frac{1}{\mathcal{Z}_{\mathrm{TL}}}\int\Diff A_{\mu}\,
    \delta\left[{\textstyle\int_{\T^4}}\diff^4x\,A_{\mu}(x)\right]
    O[A_{\mu}]\exp(-S[A_{\mu}])\,,
    \label{eq:qedtlpint}
\end{equation}
which  corresponds to restricting the integrations to the subspace of field
 configurations with a vanishing zero mode. Reusing the nomenclature from
 Ref.~\citep{Borsanyi:2014jba}, we reference the theory associated with
~\cref{eq:qedtlpint} as $\qedtl$. With the removal of the zero-mode
 redundancy, the $\qedtl$ photon two-point function is well defined:
\begin{equation}
    D_{\mu\nu}^{(\mathrm{TL})}(x-y)=
    \frac1{TL^3}\sump_{k\in\hat{\T}^4}
    \frac{\delta_{\mu\nu}}{k^2}e^{ik\cdotp(x-y)}\,,
\end{equation}
where the primed sum indicates that the $k=0$ term is excluded from the
summation.

At first sight, $\qedtl$ appears to be an acceptable solution to the zero-mode
problem, and it has in fact been used in numerous lattice QED
calculations~\citep{Duncan:1996xy,Borsanyi:2013lga,deDivitiis:2013xla,Borsanyi:2014jba}.
However, as first noticed in Ref.~\citep{Borsanyi:2014jba}, problems arise when
one tries to couple $\qedtl$ to matter fields. The source term
\begin{equation}
    S_{\text{int.}}[A_{\mu},J_{\mu}]=
    \int_{\T^4}\diff^4 x\,A_{\mu}(x)J_{\mu}(x)\,,
\end{equation}
which couples photons $(A_\mu)$ to an external current $J_\mu$, is not invariant
under shift transformations and matter therefore couples to unphysical photon
zero modes. A way out of this problem is provided by the shift-invariant
interaction term
\begin{equation}
    S_{\mathrm{TL},\text{int.}}[A_{\mu},J_{\mu}]=
    \int_{\T^4}\diff^4 x\,A_{\mu}(x) \left[ J_{\mu}(x)
    -{\textstyle\frac{1}{TL^3}\int_{\T^4}\diff^4y\,J_{\mu}(y)} \right] \,.
    \label{eq:qedtlsint}
\end{equation}
However, dealing with the photon zero mode in this way introduces a non-locality
in space and time: the field $A_{\mu}$ at a point $x$  couples to $J_{\mu}$ at
all points in spacetime. This seems to be unavoidable if one wants to completely
decouple the zero mode from the theory. In $\qedtl$, this has severe
consequences: if for instance $J_{\mu}$ is not a classical background current
but the fermionic vector current $\bar{\psi}\gamma_{\mu}\psi$, the non-locality
in time renders the definition of a bounded transfer matrix for the matter
fields impossible. As a consequence, this theory cannot be continued to a
quantum field theory in Minkowski spacetime and it has a divergent $T\to+\infty$
limit. This divergence has been shown explicitly for masses of spin $0$ and
$\frac12$ particles calculated in $\qedtl$ in Ref.~\citep{Borsanyi:2014jba}.
This problem could be circumvented by  taking the $L\to+\infty$ limit first,
ending up in a theory equivalent to QED at finite temperature which has the
correct $T\to+\infty$ (zero-temperature) limit.
\subsection{The $\qedl$ theory}
An alternative way of dealing with the zero mode which maintains locality of the
interaction term in time is provided by
\begin{equation}
    S_{\mathrm{L},\text{int.}}[A_{\mu},J_{\mu}]=
    \int_{\T^4}\diff^4 x\,A_{\mu}(x) \left[ J_{\mu}(x)
    -{\textstyle\frac{1}{L^3}\int_{\T^3}\diff^3\mathbf{y}\,
    J_{\mu}(t,\mathbf{y})} \right] \,,
    \label{eq:qedlsint}
\end{equation}
where $x=(t,\mathbf{x})$ and $\T^3$ is the 3-dimensional periodic space of
extent $L$. This term is shift invariant under the symmetry group
\begin{equation}
    A_{\mu}(x)\mapsto A_{\mu}(x)+\frac{b_{\mu}(t)}{L^3}\,,
    \label{eq:qedlshift}
\end{equation}
where $b_{\mu}(t)$ is an arbitrary smooth four-vector function of the time
coordinate with mass dimension $-2$. The modification to the current
in~\cref{eq:qedlsint} can be interpreted as placing a uniform charge (current)
density in the volume, whose effect is to restore Gauss' (Ampere's) law in a
finite volume with periodic boundary conditions~\citep{Hayakawa:2008an}. As a
result, matter is decoupled from all field configurations which are constant in
space. In an infinite volume, fields are however assumed to vanish at infinity
and so these configurations  seem unphysical yet again. Following a procedure
similar to the one laid out in the previous section,  all spatial zero modes can
be removed by introducing the shift symmetry fixing term
$\delta[\int_{\T^3}\diff^3\mathbf{x}\,A_{\mu}(t,\mathbf{x})]$. As we know from the
previous discussions of $\qedtl$, the photon action is invariant under the shift
transformation in~\cref{eq:qedlshift} only through the presence of constant
modes. Removing these modes leads to the path integral
\begin{equation}
    \braket{O}=\frac{1}{\mathcal{Z}_{\mathrm{L}}}\int\Diff A_{\mu}\,
    \delta\left[{\textstyle\int_{\T^3}}\diff^3\mathbf{x}
    \,A_{\mu}(t,\mathbf{x})\right]
    O[A_{\mu}]\exp(-S[A_{\mu}]-S_{\mathrm{L},\text{int.}}[A_{\mu},J_{\mu}])\,.
\end{equation}
This integral defines $\qedl$. It was first proposed in
Ref.~\citep{Hayakawa:2008an}, and the differences between $\qedl$ and $\qedtl$
were later discussed in Ref.~\citep{Borsanyi:2014jba}. It can be shown that the
$\qedl$ action fulfills the requirement of reflection positivity, which
therefore guarantees the existence of a well-defined transfer matrix and can be
analytically continued to a Minkowski quantum field theory. As a consequence,
when a mass gap exists, as in the case of hadronic states in QCD, observables
asymptote to their value at $T\to+\infty$ exponentially fast. As $\qedl$ will be
considered in the rest of this paper, the temporal extent is assumed to be
infinite in all discussions that will follow.

Although $\qedl$ solves problems associated with defining a transfer matrix, it
remains non-local in space. Naively, this non-locality is merely a finite-volume
effect and  all correlation functions computed in $\qedl$ are expected to
converge to those of QED in the infinite-volume limit. This is certainly true
classically: the non-local term in~\cref{eq:qedlsint} vanishes in the
infinite-volume limit. One might worry, however, that short-distance divergences
could in principle couple to volume effects through radiative corrections,
making the renormalization of $\qedl$  ambiguous. In fact, hints of subtleties
arising from an incomplete decoupling of short-distance and long-distance
effects are seen in attempting to describe the interactions of massive matter
fields with photons in $\qedl$ using a heavy-field effective theory approach, as
discussed in~\cref{sec:EFT}. Nonetheless, quantities that have been studied
to date with lattice QCD+QED calculations appear not to suffer from this
problem. Whether this will become an issue in future higher-precision
calculations, or in calculations of other quantities, remains to be determined.

At the core of the issues discussed here are the periodic boundary conditions
which allow for constant field configurations to be present. Authors of
Ref.~\citep{Lucini:2015hfa} proposed to use boundary conditions for which all
particles undergo a charge conjugation transformation at every period. In that
formulation, the photon field is antiperiodic and therefore does not have a zero
mode, allowing for the definition of a local theory in a finite volume. However,
this theory exhibits a few non-trivial features, such as the non-conservation of
electric charge and flavor quantum numbers. Realistic numerical simulations
implementing this construction are underway~\citep{Hansen:2017zly,
Campos:2017fly, Hansen:2018zre} and can establish if there are advantages to be
gained with this formulation of QED in finite volume compared with others, such
as those introduced in this section, or one in which the zero mode of the photon
is avoided by introducing a small photon mass~\citep{Endres:2015gda}. In
$\qedm$, the photon is endowed with a small mass, $m_\gamma$, that localizes the
range of the electromagnetic interactions between charged particles to volumes
within a radius of $\sim 1/m_\gamma$, screening electric charges. This
construction explicitly violates Gauss's Law without the removal of modes of the
electromagnetic field by including a mass gap into the theory. Consequently, in
volumes of spatial extent $L$, modifications to localized observables,
such as the mass of a charged particle, are exponentially insensitive to the
finite volume for $m_\gamma L\sim 4$. However, the observables depend explicitly
upon $m_\gamma$. To recover infinite-volume values of observables, an
extrapolation to $m_\gamma=0$ is required. Using (local) EFTs to cleanly
separate UV and IR physics scales, the counterterms defining the EFT  are
polynomial functions of $m_\gamma$, which can be determined by fitting to
results of lattice calculations performed in a range of volumes that satisfy
$m_\gamma L \sim 4$. The  $m_\gamma=0$ values can then be identified. The EFT
can be used, with these $m_\gamma=0$ counterterms, to make predictions. For
this technique to provide a complete quantification of uncertainties, a
hierarchy of scales between $m_\gamma$ and the lightest hadron mass $m_H$,
$m_\gamma/m_H\ll 1$, must exists to ensure that the counterterms only involve
polynomials in $m_\gamma$. This in turn requires large-volume lattice
simulations, with volumes of QCD+$\qedm$ calculations that are significantly
larger than for QCD calculations.

%% file: sections/fv.tex
In this section, we compute  the volume dependence of the $\qedl$ self-energy
functions for spins $0$ and $\frac12$ fundamental charged particles. Such
corrections are obtained for the first time in the present paper for moving
particles in a finite cubic volume with periodic boundary conditions, both on
and away from their mass shell.
\subsection{Finite-volume effects in the self-energy function}
\subsubsection{Formal definition}
\label{eq:sedef}
We consider $\qedl$, as described in the previous section, coupled to QCD
interaction through the usual gauge-covariant coupling of the Dirac quark action
to $\qedl$ and the $\SU(3)$ Yang-Mills gauge action representing the gluon
fields. In this context, we are interested in the propagation amplitude of a
given hadronic state $\ket{X}$, which can be interpolated through the 2-point
function of a given operator $\phi_X$ with the same conserved quantum numbers as
the physical state $\ket{X}$. Expanding in the electromagnetic coupling
constant, we obtain the second-order electromagnetic corrections to this 2-point
function as $D_{\mu\nu}(y_1-y_2)\,C_{\mu\nu}(x_1,y_1,x_2,y_2)$, where
\begin{equation}
  C_{\mu\nu}(x_1,y_1,x_2,y_2)=\bra{0}\mathrm{T}
  [\phi_X(x_1)J_{\mu}(y_1)J_{\nu}(y_2)\phi_X(x_2)^{\dagger}]\ket{0}\,,
  \label{eq:gen2pt}
\end{equation}
and where $J_{\mu}$ is the electromagnetic current. The associated
momentum-space correlation function $G_{\mu\nu}(p_1,k_1,k_2)$ can be defined
through
\begin{align}
  G_{\mu\nu}(p_1,k_1,k_2)=\int\diff^4x_1\int\diff^4y_1\int\diff^4y_2\,
  C_{\mu\nu}(x_1,y_1,0,y_2)
  e^{i(-p_1\cdotp x_1-k_1\cdotp y_1+k_2\cdotp y_2)}\,.
\end{align}
This function can be amputated to define the corresponding vertex function
\begin{equation}
  \Gamma_{\mu\nu}(p_1,k_1,k_2)=G(p_1)^{-1}G_{\mu\nu}(p_1,k_1,k_2)
  G(p_1+k_1-k_2)^{-1}\,,\label{eq:verfn}
\end{equation}
where $G(p)$ is the momentum-space 2-point function of field $\phi_X$,
\begin{equation}
  G(p)=\int\diff^4 x\bra{0}\mathrm{T}[\phi_X(x)\phi_X(0)^{\dagger}]\ket{0}
  e^{-ip\cdotp x}\,.
\end{equation}
From these definitions, the self-energy function of $X$ is given by
\begin{equation}
  \Sigma(p)=\int\frac{\diff^4k}{(2\pi)^4}K(k,p)\,,
\end{equation}
where $K(k,p)$ is the electromagnetic kernel,
\begin{equation}
  K(k,p)=D_{\mu\nu}(k)\Gamma_{\mu\nu}(p,k,k)\,,
\end{equation}
with $D_{\mu\nu}(k)$ the free photon propagator in a given gauge. In Euclidean
 spacetime, the on-shell self-energy can be obtained by setting $p=(i\om,\psp)$
 with $\om=\sqrt{\psp^2+m^2}$.

Now consider a hypercubic spacetime that has infinite extent in the time
direction, and is periodic in spatial directions with a finite extent $L$. In
this construction, the $\qedl$ self-energy is given by
\begin{equation}
  \Sigma^{(L)}(p)=\frac{1}{L^3}
  \sump_{\ksp\in\hat{\T}^3}\int\frac{\diff k_0}{2\pi}K(k,p)\,,
  \label{eq:SigmaL}
\end{equation}
where, as defined in~\cref{sec:qedl}, $\hat{\T}^3$ is the set
of three-vectors of the form $\frac{2\pi}{L}\nsp$ where $\nsp$ has integer
components. The $\qedl$ finite-volume effects in the self-energy are then given
by
\begin{equation}
  \Delta\Sigma(p)=\Sigma(p)-\Sigma^{(L)}(p)=
  \left(\frac{1}{L^3}\sump_{\ksp\in\hat{\T}^3}-
  \int\frac{\diff^3\ksp}{(2\pi)^3}\right)
  \int\frac{\diff k_0}{2\pi}\,K(k,p)\label{eq:fvdef}\,.
\end{equation}
By rescaling the loop 3-momentum by $\frac{2\pi}{L}$, this expression can be
written in the compact form
\begin{equation}
  \Delta\Sigma(p)=\frac{1}{L^3}\Delta_{\nsp}'\int\frac{\diff k_0}{2\pi}\,
  K((k_0,{\textstyle\frac{2\pi}{L}}\mathbf{n}),p)\,,\label{eq:fvmaster}
\end{equation}
where $\Delta_{\nsp}'$ is  defined as the the sum-integral difference
\begin{equation}
  \Delta_{\nsp}'=\sump_{\nsp\in\Z^3}-\int\diff^3\nsp\,.
  \label{eq:delta-sum}
\end{equation}
Because of the singularities in the kernel $K(k,p)$ at zero photon momentum,
$\Delta\Sigma(p)$ is expected to behave like a polynomial in $\frac{1}{L}$ at
large $L$. Moreover, it has been shown~\citep{Borsanyi:2014jba,Lubicz:2016xro}
using a generic low-energy effective representation of $\Gamma_{\mu\nu}$ that
the two first orders of this expansion are universal, \ie they do not depend on
the structure of the hadron $X$ and can be obtained in the point-like
approximation. This important fact is shown to be a direct consequence of gauge
invariance, which constrains the form of $\Gamma_{\mu\nu}$ through
Ward-Takahashi identities.

The main purpose of the present work is to discuss how to generalize
electromagnetic finite-volume calculations to virtual processes and moving
frames, and discuss the qualitative properties of the results. Therefore, we
will only consider the case of point-like particles, although all the
aspects discussed below generalize to the case of composite particles, given an
appropriate parametrization of the vertex function $\Gamma_{\mu\nu}$.
\subsubsection{Point-like spin $0$ and $\frac12$ electromagnetic self-energy functions} 
\begin{figure}[t]
  \includegraphics{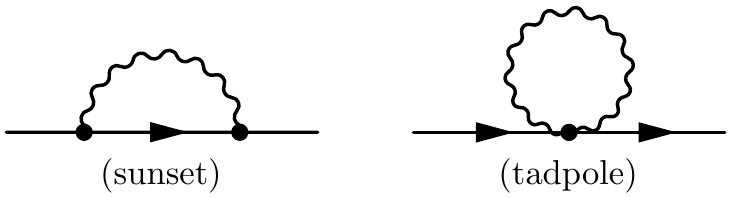}
  \caption{The self-energy of a scalar point-like particle at
  $\mathcal{O}(\alpha) $ consists of diagrams shown with one-photon and
  two-photon couplings $-q(p_1+p_2)^{\mu}$ and $-2q^2\delta_{\mu\nu}$,
  respectively, where $p_1(p_2)$ is the incoming (outgoing) particle momentum.
  The self-energy of a point-like spin-$\frac{1}{2}$ particle at $\mathcal{O}
  (\alpha)$ is given by the diagram in the left panel with the one-photon
  coupling $-q\gamma_{\mu}$, where $\gamma_{\mu}$ is an Euclidean Dirac gamma
  matrix.}
  \label{fig:feynman}
\end{figure}

In finite (infinite) Euclidean spacetime and at $\mathcal{O}(q^2)$, the
self-energy of massive point-like particles with electric charge $q$, mass $m$, and
four-momentum $p=(p_0,\mathbf{p})$ are given by sums (integrals) over the photon
momentum $k=(k_0,\ksp)$ of the kernels
\begin{align}
  K_0(k,p)&=q^2\left\{\frac{4}{k^2}-\frac{(2p-k)^2}{k^2[(p-k)^2+m^2]}\right\},
  \label{eq:K0} \\
  K_{\frac12}(k,p)&=
  q^2\left\{\frac{2i(\slashed{p}-\slashed{k})+4m}{k^2[(p-k)^2+m^2]}\right\}\,,
  \label{eq:Khalf}
\end{align}
for spins $0$ and $\frac12$, respectively (see~\cref{fig:feynman}). In what
follows, strategies to obtain the large-volume expansion of the finite-volume
effects associated with these kernels are discussed.
\subsection{Large-volume expansion}
\label{sec:lexp}
\subsubsection{General result}
\label{sec:strategy} 
It is natural to use the on-shell energy $\om=\sqrt{\mathbf{p}^2+m^2}$ as a
reference scale, in terms of which the infinite-volume limit is taken as
$L\,\om\to+\infty$. Useful dimensionless ratios with $\om$ are the
$\sigma$-\emph{ratio}
\begin{equation}
  \sigma=\frac{p^2+m^2}{\om^2} =\frac{p_0^2}{\om^2}+1\,,
\end{equation}
and the \emph{velocity}
\begin{equation}
  \vel=\frac{\mathbf{p}}{\om}\,.
\end{equation}
The $\sigma$-ratio is positive and vanishes for an on-shell external state (\ie
$p^2=-m^2$). More precisely, in Euclidean spacetime $\sigma\geq 1$ and
correlation functions can be analytically continued to $0\leq\sigma<1$, where
$\sigma=0$ is the on-shell point. The velocity has a magnitude strictly smaller
than one for massive particles. In terms of these parameters, the external
momentum can be expressed as
\begin{equation}
  p=\om(\sqrt{\sigma-1},\vel)\,,
  \label{eq:momsigv}
\end{equation}
hence any function of $p$ can be expressed as a function of $\om$, $\sigma$ and
$\vel$.

The calculation of the finite-volume effects  proceeds as follows. Consider a
kernel of the form
\begin{equation}
  K(k,p)=\frac{f(k,p)}{k^2[(p-k)^2+m^2]}\,,
\end{equation}
where the numerator $f(k,p)$ is an analytic function in $k$ and $p$. Performing
the $k_0$ integration, we obtain
\begin{equation}
  \int\frac{\diff k_0}{2\pi}\,K(k,p)
  =R_{\gamma}(\ksp,p)+R_{m}(\ksp,p)\,,
  \label{eq:mastercont}
\end{equation}
with the upper-plane residues
\begin{equation}
  R_{\gamma}(\ksp,p)=i\Res_{k_0=i|\ksp|}K(k,p)\qquad\text{and}\qquad
  R_{m}(\ksp,p)=i\Res_{k_0=p_0+i\omk}K(k,p)\,\label{eq:genres}.
\end{equation}
Now using~\cref{eq:fvmaster}, the finite-volume effects in the self-energy are
given by
\begin{equation}
  \Delta\Sigma(p)=\Delta_{\gamma}(p)+\Delta_m(p)\,,\label{eq:fvres}
\end{equation}
with
\begin{equation}
  \Delta_j(p)=\frac{1}{L^3}\Delta_{\nsp}'R_j({\textstyle\frac{2\pi}{L}\nsp},p)\,,
\end{equation}
for $j=\gamma,m$. These effects can be directly computed by studying the
behavior of the residues around $\ksp=\mathbf{0}$. In this section, only the
explicit results are presented and further details of the derivations can be
found in~\cref{sec:masterfv}.

For the on-shell momentum $p=p_{\mathrm{o.s.}}=(i\om,\psp)$, the photon-pole
finite-volume effect is given by
\begin{equation}
  \Delta_{\gamma}(p_{\mathrm{o.s.}})=
  \frac{f_0(p_{\mathrm{o.s.}})c_{2,1}(\vel)}{16\pi^2\om L}
  +\sum_{j=1}^{+\infty}\frac{\xi_{2-j,1,j}(p_{\mathrm{o.s.}})}
  {2^{4-j}\pi^{2-j}\om L^{1+j}}
  +\cdots\,,\label{eq:Dgammaos}
\end{equation}
where the ellipsis denote exponentially suppressed finite-volume effects, and
the coefficients $f_j(\hat{\ksp},p)$, $c_{j,k}(\vel)$, and $\xi_{j,k}(\psp)$ are
defined by
\begin{align}
  f((i|\ksp|,\ksp),p)&=f_0(p)
  +\sum_{j=1}^{+\infty}f_j(\hat{\ksp},p)|\ksp|^j\,,\label{eq:fexp}\\
  c_{j,k}(\vel)&=\Delta_{\nsp}'\left[\frac{1}{|\nsp|^{j}(1-\vdnu)^{k}}\right]\,,
  \label{eq:cjkdef}\\
  \xi_{j,k,l}(p)&=\Delta_{\nsp}'
  \left[\frac{f_{l}(\hat{\nsp},p)}{|\nsp|^{j}(1-\vdnu)^{k}}\right]\,.
\end{align}
The coefficient $c_{j,k}(\vel)$ that drives the leading-order correction is
particularly important as it appears systematically in perturbative calculations
of QED finite-size effects. The properties and evaluation of these numbers are
studied in detail in~\cref{sec:fvcoef}. We also define the rest-frame
coefficients
\begin{equation}
  c_j=c_{j,k}(\mathbf{0})=\Delta_{\nsp}'\frac{1}{|\nsp|^{j}}\,.
\end{equation}
These coefficients can be seen as particular values of the generalized zeta function from Ref.~\citep{Luscher:1986pf}, \ie $c_j=Z_{00}(\frac{j}{2},\mathbf{0})$, and have the known values
\begin{equation}
  c_2=\pi c_1,\qquad c_1=-2.83729748\dots,\qquad\text{and}\qquad c_0=-1\,.
\end{equation}
In the off-shell case, one obtains
\begin{equation}
  \Delta_{\gamma}(p)=
  \frac{f_{0}(p)c_1}{4\pi\sigma\om^2L^2}
  +\left[
  -\frac{i\sqrt{\sigma-1}\,f_{0}(p)}{\sigma^2\om^3}+
  \frac{\xi_{0,0,1}(p)}{2\sigma\om^2}
  \right]\frac{1}{L^3}+\bigo\left(\frac{1}{L^4}\right)\,.
  \label{eq:Dgamma}
\end{equation}
We observe that in this case the absence of the on-shell singularity pushes the
finite-volume effects to $\bigo(\frac{1}{L^2})$. The charged-particle effect
$\Delta_m(p)$ is $\bigo(\frac{1}{L^3})$ independently of on-shell conditions and
it is given by
\begin{equation}
  \Delta_m(p)
  =-\frac{r_m(\mathbf{0},p)}{L^3}+\cdots=
  -\frac{f(((i+\sqrt{\sigma-1})\om,\mathbf{0}),p)}
  {2(i+\sqrt{\sigma-1})^2\om^3L^3}+\cdots\,.
  \label{eq:Dm}
\end{equation}
\subsubsection{Spin-$0$ self-energy}
The strategy described in the previous section can be applied to the kernel
in~\cref{eq:K0}. In this case, the function $f$ is given by
\begin{equation}
  f(k,p)=q^2(3k^2-4p\cdot k+4m^2)\,,
\end{equation}
and the coefficients $f_j$ defined in~\cref{eq:fexp} are
\begin{equation}
  f_0(p)=4q^2m^2,\qquad f_1(\hat{\ksp},p)=-4q^2[ip_0+\om(\vdku)]\,,
\end{equation}
and $f_j=0$ for $j>1$. Considering~\cref{eq:Dgammaos,eq:Dgamma}, the only
required $\xi_{j,k,l}(p)$ coefficients are
\begin{equation}
  \xi_{1,1,1}(p)=4q^2\om c_1\qquad\text{and}\qquad
  \xi_{0,0,1}(p)=-4iq^2p_0c_0=4iq^2p_0\,.
\end{equation}

Substituting relevant functions in~\cref{eq:Dgammaos}, the on-shell
finite-volume effects from the photon pole are given by
\begin{align}
  \Delta_{\gamma}(p_{\mathrm{o.s.}})&=q^2\left[\frac{m^2c_{2,1}(\vel)}{4\pi^2\om L}
  +\frac{c_1}{2\pi L^2}+\cdots\right]\\
  &=m^2q^2\left[\frac{1}{\gv}\frac{c_{2,1}(\vel)}{4\pi^2\mu}  
  +\frac{c_1}{2\pi\mu^2}+\cdots\right]\,,
\end{align}
where $\mu=mL$, and $\gv=(1-|\mathbf{v}|^2)^{-1/2}$ is the usual Lorentz
contraction factor. For the scalar-particle pole, the effects are 
\begin{equation}
  \Delta_m(p_{\mathrm{o.s.}})=m^2q^2\left[
    \frac{1}{\gv^3}-\frac{1}{\gv}\right]
    \frac{1}{2\mu^3}+\cdots\,.
    \label{eq:photonforscalar}
\end{equation}
on shell. Putting everything together, the finite-volume effects on the self
energy of a moving on-shell  spin-$0$ particle are
\begin{equation}
  \Delta\omega_0(\psp)^2=\Delta\Sigma(p_{\mathrm{o.s.}})=
  m^2q^2\left\{
  \frac{1}{\gv}\frac{c_{2,1}(\vel)}{4\pi^2\mu}+\frac{c_1}{2\pi\mu^2}+
  \left[
    \frac{1}{\gv^3}-\frac{1}{\gv}\right]
  \frac{1}{2\mu^3}+\cdots\right\}\,,\label{eq:os0}
\end{equation}

In the off-shell case, using~\cref{eq:Dgammaos,eq:Dm}, the photon-pole and
particle-pole effects are
\begin{align}
  \Delta_{\gamma}(p)
  =m^2q^2\left\{\frac{1}{\gv^2}\frac{c_1}{\pi\sigma\mu^2}
 +\left[\frac{2i\sqrt{\sigma-1}}{\gv\sigma}
 -\frac{4i\sqrt{\sigma-1}}{\gv^3\sigma^2}\right]\frac{1}{\mu^3}
 +\bigo\left(\frac{1}{\mu^4}\right)\right\}\,,
\end{align}
and
\begin{equation}
  \Delta_m(p)=m^2q^2\left[\frac{\sigma-4-4i\sqrt{\sigma-1}}
  {2\sigma\gv}
  -\frac{2}{(i+\sqrt{\sigma-1})^2\gv^3}\right]\frac{1}{\mu^3}
  +\cdots,
\end{equation}
respectively. Note that this pole contribution has no on-shell singularities,
and the on-shell effects are obtained by taking the limit $\sigma\to 0$,
consistent with~\cref{eq:photonforscalar}. Finally, the finite-volume effect in
the off-shell self-energy of a moving spin-0 particle is 
\begin{equation}
  \Delta\Sigma_0(p)=m^2q^2
  \left\{\frac{1}{\gv^2}\frac{c_1}{\pi\sigma\mu^2}+
  \left[\left(\frac{4}{\sigma^2}-\frac{2}{\sigma}\right)
  \frac{1}{\gv^3}+\left(\frac{1}{2}-\frac{2}{\sigma}\right)
  \frac{1}{\gv}\right]\frac{1}{\mu^3}+\bigo\left(\frac{1}{\mu^4}\right)\right\}\,.
  \label{eq:offs0}
\end{equation}
\subsubsection{Spin $\frac12$ self-energy}
In this case, the kernel~\cref{eq:Khalf} leads to
\begin{equation}
  f(k,p)=q^2[2i(\slashed{p}-\slashed{k})+4m]\,,
\end{equation}
and the coefficients $f_j$ defined by~\cref{eq:fexp} are
\begin{equation}
  f_0(p)=q^2(2i\slashed{p}+4m)\,,\qquad f_1(p,\hat{\ksp})=
  -2iq^2(i\gamma_0+\hat{\mathbf{k}}\cdot\bm{\gamma})\,,
\end{equation}
and $f_j=0$ for $j>1$. The only required $\xi_{j,k,l}(p)$ coefficients are
\begin{equation}
  \xi_{1,1,1}(p)=-2q^2i \Delta'_{\mathbf{n}}\frac{i\gamma_0
  +\bm{\gamma}\cdot\mathbf{n}}{|\mathbf{n}|(1-\mathbf{v}\cdot \mathbf{n})}
  \qquad\text{and}\qquad
  \xi_{0,0,1}(p)=-2q^2\gamma_0\,.
\end{equation}

The on-shell condition is achieved through substitutions
\begin{align}
\gamma_0 \mapsto \frac{\omega(\mathbf{p})}{m}=\gv,
\\
\bm{\gamma} \mapsto -i\frac{\mathbf{p}}{m}=i\gv|\mathbf{v}|,
\end{align}
which lead to the desired on-shell condition $\slashed{p}=im$ in Euclidean
spacetime. Next, using~\cref{eq:Dgammaos}, the on-shell finite-volume effects
from the photon pole are given by
\begin{align}
  \Delta_{\gamma}(p_{\mathrm{o.s.}})&=
  q^2m\left[\frac{1}{\gv}\frac{c_{2,1}(\vel)}{8\pi^2\mu}  
  +\frac{c_1}{4\pi\mu^2}+\cdots\right]\,.
\end{align}
From the fermion pole on the other hand, one obtains
\begin{align}
  \Delta_{m}(p_{\mathrm{o.s.}})&=
  q^2m\left[\frac{1}{4\gv^3\mu^3}+\frac{1}{2\gv\mu^3}+\cdots\right]\,
\end{align}
on shell. Finally, the full on-shell finite-volume corrections are
\begin{equation}
  \Delta\omega_{\frac12}(\psp)=q^2m\left\{
  \frac{1}{\gv}\frac{c_{2,1}(\vel)}{8\pi^2\mu}+\frac{c_1}{4\pi\mu^2}+
  \left[\frac{2}{\gv}+\frac{1}{\gv^3}\right]\frac{1}{4\mu^3}+\cdots\right\}\,.
  \label{eq:oshalf}
\end{equation}

Finally, straightforward algebra leads the full finite-volume effects in the
off-shell self-energy of spin-$1/2$ particles,
\begin{equation}
\Delta\Sigma_{\frac12}(p)=q^2\left\{\frac{(i\slashed{p}+2m)c_1}{2\pi\sigma\gv^2
  \mu^2}+
  \left[-\frac{i\om(\vel\cdotp\boldsymbol{\gamma})+2m}{\sigma}
  +\frac{2i\slashed{p}+4m}{\sigma^2}\right] \frac{1}{\gv^3\mu^3}
  +\mathcal{O}\left(\frac{1}{\mu^4}\right)
  \right\}\,.\label{eq:offshalf}
\end{equation}
\subsubsection{Universality of the on-shell corrections}
The finite-volume corrections to the energy of charged spin-$0$ and
spin-$\frac12$ particles, \ie\cref{eq:os0,eq:oshalf}, when evaluated in their
rest frames, correctly reproduce the results found in
Refs.~\citep{deDivitiis:2013xla, Borsanyi:2014jba}. The result from
Ref.~\citep{Davoudi:2014qua} has a different $\mathcal{O}(\frac{1}{L^3})$ term
in the spin $\frac12$ case, which is due to subtleties in the construction of
non-relativistic QED with a non-local theory as $\qedl$. This issue was first
commented in Ref.~\citep{Fodor:2015pna} and is investigated with more details
in~\cref{sec:EFT} of the present paper.

An important result derived in Ref.~\citep{Borsanyi:2014jba}, which was
developed and extended in Ref.~\citep{Lubicz:2016xro}, is the universality of
the $\mathcal{O}(\frac{1}{L})$ and $\mathcal{O}(\frac{1}{L^2})$ finite-volume
corrections to the mass. The statement goes as follows: even in the case where
the particle is not elementary, but rather a composite bound state of the strong
interaction, the $\mathcal{O}(\frac{1}{L})$ and $\mathcal{O}(\frac{1}{L^2})$
electromagnetic finite-size corrections to the mass are identical to the case of
a point particle. This property is a consequence of gauge invariance, which
through Ward identities strongly constrains the form of the on-shell vertex
function $\Gamma_{\mu\nu}(p,k,k)$, defined in~\cref{eq:sedef}, in the
soft-photon limit $k^2\to 0$. More precisely, the leading singularities in $k$
in the electromagnetic kernel $K(k,p)$, responsible for the leading power
corrections in $\frac{1}{L}$, are independent of the particle structure. The
universality argument was derived in
Refs.~\citep{Borsanyi:2014jba,Lubicz:2016xro} with arbitrary kinematics, and is
naturally applicable to the new results presented here
in~\cref{eq:os0,eq:oshalf} for the on-shell self-energy in a moving frame.
Moreover, identically to the rest frame case, one can notice that the
finite-volume effects on the self-energy are independent of the spin up to
$\mathcal{O}(\frac{1}{L^3})$ effects:
\begin{equation}
  2m\Delta\Sigma_{\frac12}(p)=
  \Sigma_{0}(p)^2+\mathcal{O}({\textstyle\frac{1}{L^3}})\,,
\end{equation}
once the conversion between mass and squared-mass is made at leading order in
$q^2$. As is seen in~\cref{eq:offs0,eq:offshalf}, such universality does not
extend to the off-shell results.
\subsection{Finite-volume coefficients}
\label{sec:fvcoef}
In this section, the properties of the coefficients $c_{j,k}(\vel)$ that drive
the large volume expansion will be discussed.
\subsubsection{Rest-frame coefficients}
\begin{figure}[t]
  \includegraphics{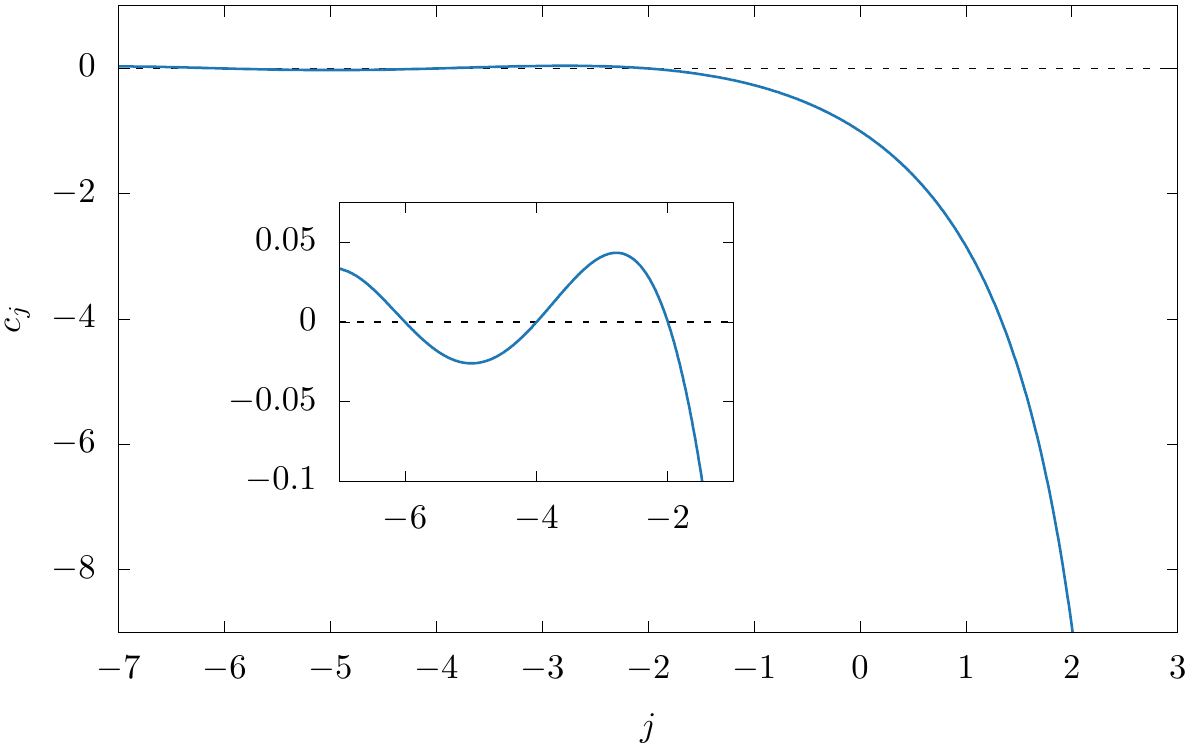}
  \caption{The rest-frame finite-volume coefficients $c_j$ as a function of
  $j$. The inset panel is a zoom on the small oscillations in the 
  $-7\leq j\leq -1$ region, with zeros on even negative integers.}
  \label{fig:cj}
\end{figure}

Consider the rest-frame coefficients
\begin{equation}
  c_j=\Delta_{\nsp}'\frac{1}{|\nsp|^{j}}.
\end{equation}
This number is only well defined for $j<3$. For $j\geq 3$ the function
$|\nsp|^{-j}$ is no longer integrable around $\nsp=\mathbf{0}$. This singularity
is physically related to the presence of electromagnetic infrared divergences in
the infinite-volume amplitude. One such case has been studied in
Ref.~\citep{Lubicz:2016xro}, and in the present work we will only consider
infrared finite quantities. Using the fact that $|\nsp|^{j}$ is polynomial in the
components of $\nsp$ for even integers $j$, one obtains
\begin{equation}
  c_0=-1\qquad\text{and}\qquad c_{-j}=0\quad\text{for }j\text{ even}\,.
\end{equation}
Also, for $j>0$, the Poisson summation formula gives the interesting reflection
formula
\begin{equation}
  c_j=\pi^{j-\frac{3}{2}}\frac{\Gamma\left(\frac{3-j}{2}\right)}
  {\Gamma\left(\frac{j}{2}\right)}c_{3-j}\,,
\end{equation}
which is a known property of these sums~\citep{Borwein:2013}. This
relation gives the useful identity~\citep{Davoudi:2014qua}
\begin{equation}
  c_2=\pi c_1\,,
\end{equation}
and determines the divergent asymptotic behavior of $c_j$ for $j\to 3$,
\begin{equation}
  c_j\underset{j\to 3^-}{\sim}\frac{4\pi}{j-3}\,.
\end{equation}
Naively, the numerical evaluation of $c_j$ is not straightforward as it
emerges from the cancellation of a divergent series with a divergent integral.
For this work, we developed an accelerated evaluation of $c_j$ with a doubly
exponential rate of convergence. The method is presented in~\cref{sec:cj}, and
was used in the numerical applications that follow. Finally, we present the 
values of $c_j$ as function of $j$ in~\cref{fig:cj} and give useful 
values in~\cref{tab:cj}.
\begin{table}[t]
  \begin{tabular}{|c|c|}
      \hline
      $j$ & $c_j$\\
      \hline
      $-5$  & $-0.02587$         \\
      $-3$  & $0.04118$         \\
      $-1$  & $-0.26660$         \\
      $0$   & $-1$         \\
      $1$   & $-2.83730$\\
      $2$   & $-8.91363$\\
      \hline
  \end{tabular}
  \caption{Values of selected zero-velocity finite-volume coefficients.}
  \label{tab:cj}
\end{table}
\subsubsection{Moving-frame coefficients and rotational symmetry breaking effects}
\begin{figure}[t]
  \includegraphics{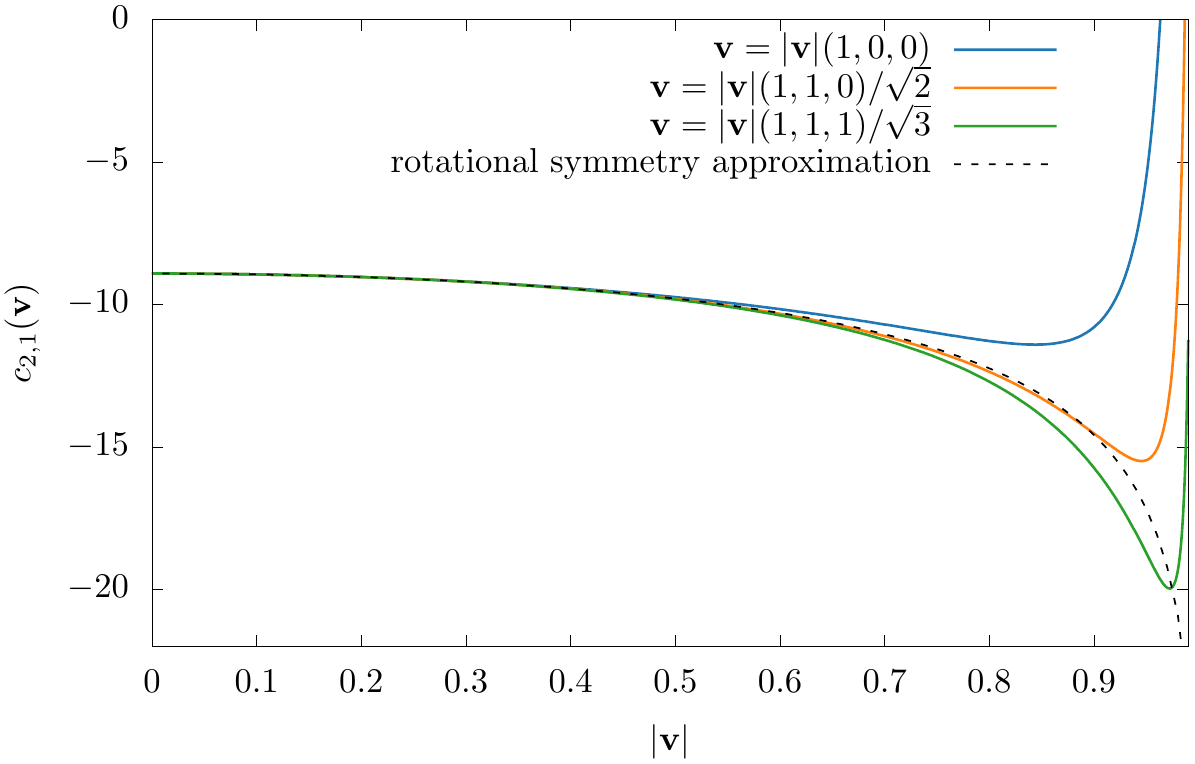}
  \caption{The finite-volume coefficient $c_{2,1}(\vel)$ as a function of the
  velocity norm for typical orientations of the velocity compared to the
  rotational symmetry approximation given in~\cref{eq:c21rot}. The values of
  $c_{2,1}(\vel)$ displayed here have been computed numerically using the
  techniques described in~\cref{sec:cj}.}
  \label{fig:c21}
\end{figure}

In a finite cubic volume, the rotational symmetry group is broken down to the
cubic symmetry group. Therefore, at non-zero velocity, finite-volume effects
will depend on the direction of the vector $\vel$. Upon inspecting the results
of the previous section, it becomes clear that the rotational symmetry breaking
effects will be encoded in the dependence of the $c_{2,1}(\vel)$ coefficient on
the velocity direction. This can be explored in more detail by means of a
spherical harmonic analysis of the angular dependence of $c_{2,1}(\vel)$.
Consider the spherical expansion
\begin{equation}
  \frac{1}{|\nsp|^j(1-\vdnu)^k}=\frac{1}{|\nsp|^j}
  \sum_{l=0}^{+\infty}\sum_{m=-l}^l
  a_{klm}(\vel)Y_{lm}(\theta_{\nsp},\phi_{\nsp})\,,
  \label{eq:ylmexp}
\end{equation}
where $Y_{lm}$ is the normalized spherical harmonic
\begin{equation}
  Y_{lm}(\theta,\phi)=\sqrt{\frac{2l+1}{4\pi}}
  \sqrt{\frac{(l-m)!}{(l+m)!}}P_{lm}[\cos(\theta)]e^{im\phi}\,,
\end{equation}
with $P_{lm}$ the associated Legendre polynomial. Moreover, $\theta_{\nsp}$
and $\phi_{\nsp}$ are the angular spherical coordinates of $\nsp$,
\begin{equation}
  \usp=\left(\sin(\theta_{\nsp})\cos(\phi_{\nsp}),
  \sin(\theta_{\nsp})\sin(\phi_{\nsp}),
  \cos(\theta_{\nsp})\right)\,.
\end{equation}
Rotational symmetry requires that  the integrals over $\nsp$ of the terms
in~\cref{eq:ylmexp} vanish except for the $l=0$ term, allowing the
$c_{j,k}(\vel)$ coefficients in~\cref{eq:cjkdef} to be written as
\begin{equation}
  c_{j,k}(\vel)=A_k(|\vel|)c_j
  +\sum_{l=1}^{+\infty}\sum_{m=-l}^la_{klm}(\vel)y_{jlm}\,,
  \label{eq:cjklmexp}
\end{equation}
where
\begin{align}
  A_k(\beta)&=
 \frac12\int_{-1}^{1}\frac{\diff x}
  {(1-\beta x)^k}
 =\frac{1}{2\beta(k-1)}
  \left[
  \left(\frac{1}{1-\beta}\right)^{k-1}-
  \left(\frac{1}{1+\beta}\right)^{k-1}\right] ,
\end{align}
and
\begin{align}  
  \label{eq:Adef}  y_{jlm}&=\Delta_{\nsp}'
  [|\nsp|^{-j}Y_{lm}(\theta_{\nsp},\phi_{\nsp})]=
  \sump_{\nsp}\frac{Y_{lm}(\theta_{\nsp},\phi_{\nsp})}{|\nsp|^j}\,.
\end{align}
The coefficients $y_{jlm}$ are given in terms of the generalized zeta function
from Ref.~\citep{Luscher:1986pf} through the relation 
\begin{equation}
  y_{jlm}=\sqrt{\frac{4\pi}{2l+1}}Z_{lm}\left(\frac{j-l}{2},\mathbf{0}\right)\,.
\end{equation}
The expression in~\cref{eq:cjklmexp} is the main result of this section, and
shows that the $c_{j,k}(\vel)$ coefficients, up to rotational symmetry breaking
effects, are proportional to $c_j$ by a known factor depending only on the
magnitude of the velocity $\mathbf{v}$. Rotational-symmetry breaking effects
enter through the higher multipole contributions $a_{klm}(\vel)$. The function
$A_k(\beta)$ has the limits
\begin{equation}
  A_k(0)=1\qquad\text{and}\qquad A_1(\beta)=\frac{\arctanh(\beta)}{\beta}\,.
\end{equation}
In the case that is relevant to QED finite-volume corrections, the
rotational symmetry approximation $\bar{c}_{2,1}(|\vel|)$ of $c_{2,1}(\vel)$ is
given by
\begin{equation}
  \bar{c}_{2,1}(|\vel|)=\frac{\pi c_1}{|\vel|}\arctanh(|\vel|)\,,
  \label{eq:c21rot}
\end{equation}
where the presence of the rapidity $\arctanh(|\vel|)$ is noted.
In~\cref{fig:c21}, the exact value of $c_{2,1}(\vel)$ is compared to
$\bar{c}_{2,1}(|\vel|)$ for sample velocity orientations.
The rotational symmetry approximation appears to be very good up to velocities
$|\vel|\sim 0.5$. At ultra-relativistic velocities, the rotational symmetry
breaking effects dominate. As it is proven in details in~\cref{sec:aklm},
$a_{klm}(\vel)=\mathcal{O}(|\vel|^l)$. Using this property,~\cref{eq:ylmexp} can
be interpreted as a power expansion in $|\vel|$, explaining the
ultra-relativistic behavior.

%% file: sections/improvement.tex
$\qedl$ is a minimal choice to implement QED in a finite volume in which photon
zero-mode singularities are regulated by introducing a particular form of
non-locality in space while preserving locality in time. Non-minimal choices are
possible as well and lead to different approaches to the infinite-volume limit.
Such extra non-localities can be tuned to remove or suppress finite-volume
effects. This approach has similarities to Symanzik's improvement program that
subtracts discretization effects in lattice gauge theories. Given this, we call
the method detailed below \emph{infrared improvement}.

Although knowledge of the analytic form of leading finite-volume effects in
$\qedl$ in principle suffices to subtract them out in obtaining the
infinite-volume values of quantities, it is still advantageous to carry out
numerical calculations in an improved scheme. Consider a situation in which the
finite-volume value is significantly different than the infinite-volume value,
which can be the case in relatively small volumes. Then the computational
resources required to accurately perform the required subtraction will be
significant, prohibiting precision calculations of some quantities in $\qedl$.
As shown below, a relatively general infrared improvement of $\qedl$ leads to
mass corrections of the order of sub-percent level even at small volumes, which
would be comparable to or smaller than other systematics in most
state-of-the-art numerical calculations. Similar limitations are encountered in
studies of moments of parton distribution functions of hadrons with lattice QCD,
where using conventional methods, contributions from lower-dimension operators
dominate over the continuum-limit contributions, requiring new ideas that
implement an improvement procedure in such calculations, see \eg
Refs.~\citep{Davoudi:2012ya, Detmold:2005gg, Monahan:2015lha}. Similar ideas
have been suggested in taking advantage of numerical simulations with multiple
center-of-mass boosts or boundary conditions to find optimal combination of
quantities that suppress finite-volume effects in cases where the target
quantity is small compared with other scales in the system, such as the deuteron
binding energy or the S-wave/D-wave mixing in the isosinglet two-nucleon
system~\citep{Davoudi:2011md,Briceno:2013bda,Briceno:2013hya}.

An additional motivation for an infrared-improved $\qedl$ concerns studies of
systems with multiple charged hadrons. For example, as is demonstrated in
Ref.~\citep{Beane:2014qha}, the power-law corrections to the mass of charged
particles modify the kinematics of $2 \to 2$ scattering processes, requiring
keeping track of this change in subsequent calculations. Starting out with the
incoming and outgoing hadrons that are already close to their infinite-volume
mass simplifies the formalism that extracts scattering amplitudes from energy
spectra. Additionally, the relatively general improvement scheme introduced
in~\cref{subsec:cumulative} suggests that such a single-body improvement may
lead to an improvement in finite-volume corrections in two and multi-hadron
observables, a statement that will be investigated in future studies.
\subsection{General concept}
Consider the $\qedl$ action described in~\cref{sec:qedl}, written in momentum
space,
\begin{equation}
    S_L[\hat{A}_{\mu}]=\frac1{2L^3}\int\frac{\diff k_0}{2\pi}
    \sump_{\ksp\in\hat{\T}^3}\hat{A}_{\mu}(k)^*\Omega_{\mu\nu}(k)\hat{A}_{\nu}(k)\,,
\end{equation}
where the decoupled spatial zero-mode is removed and the kernel
$\Omega_{\mu\nu}(k)$ is given by
\begin{equation}
    \Omega_{\mu\nu}(k)=\delta_{\mu\nu}k^2-k_{\mu}k_{\nu}\,.
\end{equation}
Note that the gauge has not yet been fixed. In momentum space, gauge invariance
can be summarized by the identity
\begin{equation}
    k_{\mu}\Omega_{\mu\nu}(k)=0\,,
    \label{eq:ktrans}
\end{equation}
so the tensor $\Omega_{\mu\nu}(k)$ is transverse for any $k$. Now let us define
the \emph{infrared-improved action} through
\begin{equation}
    S_{L,w}[\hat{A}_{\mu}]=\frac1{2L^3}\int\frac{\diff k_0}{2\pi}
    \sump_{\ksp\in\hat{\T}^3}
    \frac{\hat{A}_{\mu}(k)^*\Omega_{\mu\nu}(k)\hat{A}_{\nu}(k)}{1+w_{|\nsp|^2}}\,,
    \label{eq:impaction}
\end{equation}
where $\nsp=\frac{L}{2\pi}\ksp$ and the $w_{|\nsp|^2}$ are real coefficients
which are non-zero only for a finite number of values of $|\mathbf{n}|$. Because
of this property, the contributions from the $w_{|\nsp|^2}$
vanish in the infinite-volume limit. To preserve the positivity of the action,
an additional constraint, $w_{|\nsp|^2}>-1$ is placed on the coefficients for
any $\nsp$. As the only effect of introducing $w_{|\nsp|^2}$ coefficients is to
reweight the action kernel,~\cref{eq:ktrans} still holds and the theory
remains gauge invariant. Gauge fixing and integrating out the redundant gauge
degree of freedom results in a kernel that is an invertible matrix
$\bar{\Omega}_{\mu\nu}(k)$, \eg $\delta_{\mu\nu}k^2$ in Feynman gauge.
In the Euclidean quantum field theory associated with~\cref{eq:impaction},
the momentum-space photon propagator is 
\begin{align}
    \hat{D}_{\mu\nu}^{(L,w)}(k)&=
    (1+w_{|\nsp|^2})\bar{\Omega}^{-1}_{\mu\nu}(k)
    =(1+w_{|\nsp|^2})\hat{D}_{\mu\nu}^{(L)}(k)\,,
\end{align}
where $\hat{D}_{\mu\nu}^{(L)}(k)$ is the $\qedl$ photon propagator. The weight
functions $w_{|\nsp|^2}$ modify the residue of the photon propagator near
its pole, and as was demonstrated in~\cref{sec:fv}, the coefficients of the
large-volume expansion depend on this residue. The strategy of the infrared
improvement is to tune a finite number of  $w_{|\nsp|^2}$ to reduce the
finite-volume effects.

More explicitly, because of its discrete and finite nature, multiplying the
photon propagator by $1+w_{|\nsp|^2}$ does not change the singularity structure
of the contour integral in~\cref{eq:mastercont}. Therefore, the general
formulas~\cref{eq:Dgamma,eq:Dgammaos,eq:Dm} still holds by replacing the
coefficients $c_{j,k}(\vel)$ and $\xi_{j,k,l}(p)$ by
\begin{align}
  c_{j,k}^{(w)}(\vel)&=c_{j,k}(\vel)+\sump_{\nsp}
  \frac{w_{|\nsp|^2}}{|\nsp|^{j}(1-\vdnu)^{k}}\,,\label{eq:impcjk}\\
  \xi_{j,k,l}^{(w)}(p)&=\xi_{j,k,l}(p)
  +\sump_{\nsp}\frac{w_{|\nsp|^2}f_{l}(\hat{\nsp},p)}{|\nsp|^{j}(1-\vdnu)^{k}}\,,
\end{align}
respectively. We now discuss in detail different strategies to tune the weights
$w_{|\nsp|^2}$ for the self-energy functions.
\subsection{Infrared improvement of the self-energy}
\label{sec:seirimp}
As an example, let us consider the case of the on-shell scalar self-energy, for
which the finite-volume contribution was derived in~\cref{sec:lexp}. From what
was described in the previous section, in the infrared-improved theory one
obtains
\begin{equation}
  \Delta\omega_0(\psp)^2=m^2q^2\left\{
  \frac{1}{\gv}\frac{c_{2,1}^{(w)}(\vel)}{4\pi^2\mu}
  +\frac{c_1^{(w)}}{2\pi\mu^2}
  -\left[\frac{1}{\gv}-\frac{1}{\gv^3}\right]
  \frac{c_0^{(w)}}{2\mu^3}+\cdots\right\}\,.\label{eq:os0imp}
\end{equation}
It is useful to define the finite sum,
\begin{equation}
    \sigma_{k,N}(\vel) \equiv \sum_{|\nsp|^2=N}(1-\vdnu)^{-k}\,,
\label{eq:sigma-def}    
\end{equation}
from which the coefficients $c_{j,k}^{(w)}(\vel)$ become
\begin{equation}
    c_{j,k}^{(w)}(\vel)=c_{j,k}(\vel)+
    \sum_{N=1}^{+\infty}\frac{w_N}{N^{\frac{j}{2}}}\sigma_{k,N}(\vel)\, .
    \label{eq:impcjk2}
\end{equation}
Note that the sum in~\cref{eq:sigma-def} only runs over integer vectors
$\mathbf{n}$ with length $N$. For zero velocity, $\sigma_{k,N}(\vel)$ becomes
equal to $r_3(N)$, the number of integer solutions to the equation
$x^2+y^2+z^2=N$. The values of this function which are relevant here are
\begin{equation}
    r_3(1)=6,\qquad\text{and}\qquad r_3(2)=12\,.
\end{equation}

It is interesting to  consider the possibility of completely canceling
finite-volume contributions up to a given order in the $\frac{1}{L}$ expansion.
Unfortunately, this is not always possible to achieve and does not have a clear
qualitative benefit in typical physical scenarios, as is shown below for the
case of the mass. One may therefore explore the possibility of
approximately canceling the sum of several orders in the $\frac{1}{L}$ expansion
in a given reference volume. For the  finite-volume corrections to a charged
particle mass, we find that this strategy is generally more feasible and allows
for a  reduction in these effects below the percent level for volumes typically
used in lattice QCD+QED calculations.
\subsubsection{$\bigo(\frac{1}{L})$ and $\bigo(\frac{1}{L^2})$ improvements}
To completely remove the $\bigo(\frac{1}{L})$ finite-volume effects given in
~\cref{eq:os0imp,eq:impcjk2}, one needs to solve the condition
$c_{j,k}^{(w)}(\vel)=0$. A minimal choice that satisfies this is
\begin{equation}
    w_1=-\frac{c_{2,1}(\vel)}{\sigma_{1,1}(\vel)},
    \qquad\text{or}\qquad
    w_1=-\frac{\pi}{6}c_1\text{~~for~~} \vel=0\,,
\end{equation}
and $w_N=0$ for all $N>1$. This weight function modifies the
$\bigo(\frac{1}{L^2})$ and $\bigo(\frac{1}{L^3})$ coefficients to become
\begin{equation}
    c_1^{(w)}=c_1-6\frac{c_{2,1}(\vel)}{\sigma_{1,1}(\vel)},
    \qquad\text{and}\qquad
    c_0^{(w)}=-1-6\frac{c_{2,1}(\vel)}{\sigma_{1,1}(\vel)}\,.
\end{equation}
Numerical values for the $w_{|\nsp|^2}$ and $c_j^{(w)}$ presented in this
section, and in the following, are summarized in~\cref{tab:ircoef}.

The $\bigo(\frac{1}{L^2})$-improved coefficients are given by
~\cref{eq:impcjk2} through the linear system
\begin{align}
    c_{2,1}(\vel)+w_1\ \sigma_{1,1}(\vel)+\frac12\ w_2\ \sigma_{1,2}(\vel)&=0\,,\\
    c_1+6w_1+6\sqrt{2}w_2&=0\,,
\end{align}
which  gives
\begin{align}
    w_1&=\frac{c_1\sigma_{1,2}(\vel)-12\sqrt{2}c_{2,1}(\vel)}
    {6[2\sqrt{2}\sigma_{1,1}(\vel)-\sigma_{1,2}(\vel)]}\\
    \text{or}\qquad
    w_1&=-\frac16(1+\sqrt{2})(\sqrt{2}\pi-1)c_1=3.93053406\dots
    \text{~~for~~} \vel=0\,,\\
    w_2&=\frac{6c_{2,1}(\vel)-c_1\sigma_{1,1}(\vel)}
    {3[2\sqrt{2}\sigma_{1,1}(\vel)-\sigma_{1,2}(\vel)]}\\
    \text{or}\qquad
    w_2&=\frac16(1+\sqrt{2})(\pi-1)c_1=-2.44492857\dots
    \text{~~for~~} \vel=0\,,
\end{align}
where the least number of coefficients that allow a full cancellation of
$\bigo(\frac{1}{L})$ and $\bigo(\frac{1}{L^2})$ effects are considered. In the
rest frame we obtain $w_2\leq -1$ which violates the positivity of the action in
~\cref{eq:impaction}. Similar occurrences of the same issue arising in other
situations were found in attempting to exactly cancel the finite-volume effect
up to a given order, warranting investigations into  other forms of constraints
on the weight factors $w_{|\nsp|^2}$. Note that at this order, the finite-volume
corrections to the mass of  spin-0 and spin-$\frac{1}{2}$ particles are the same
and the same weight factors apply to both  cases.

\begin{figure}[p]
    \includegraphics{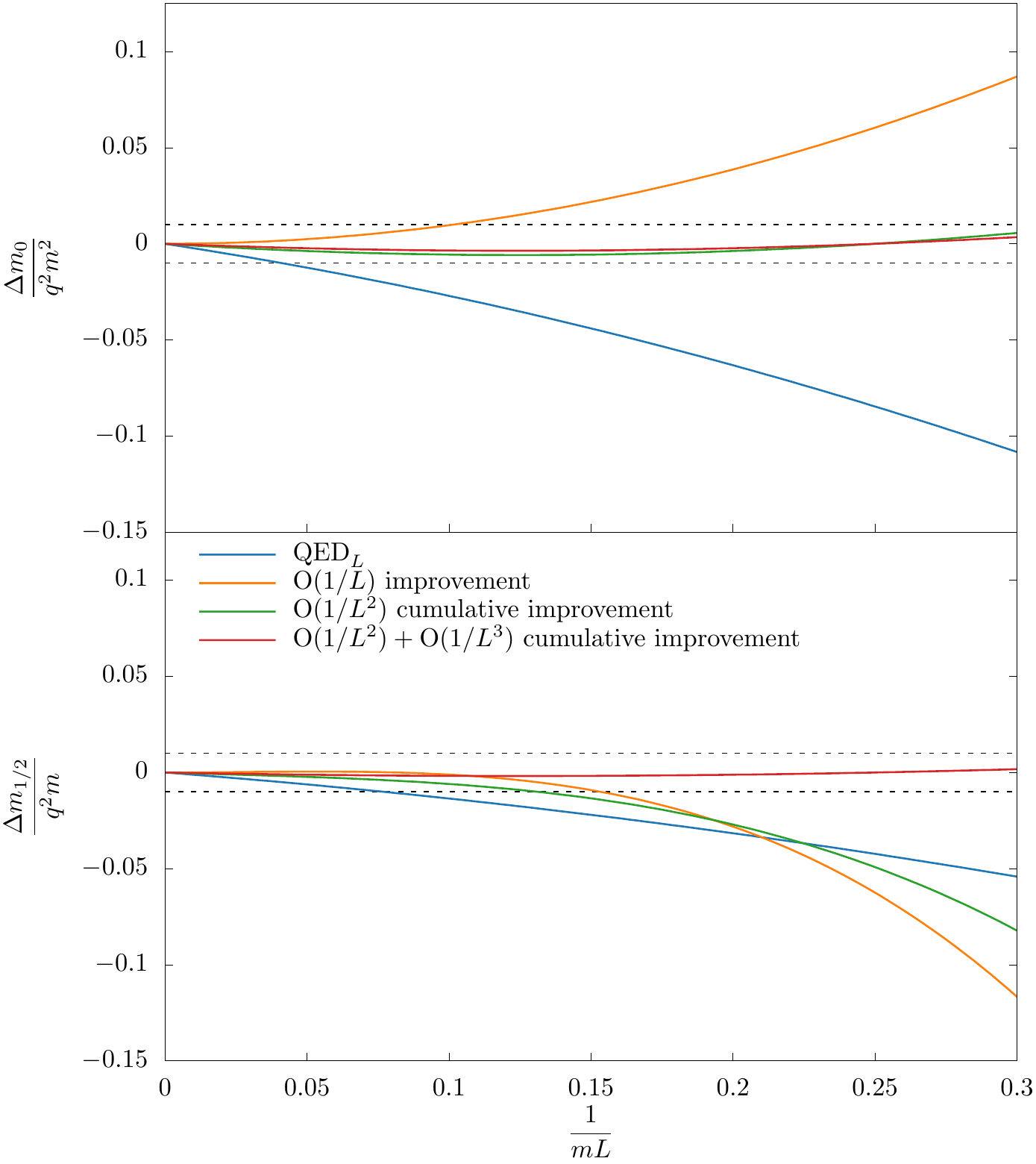}
    \caption{Relative finite-volume contributions to charged-particle masses in
    $\qedl$, and the improved versions described in~\cref{sec:seirimp}. The
    upper panel shows corrections to the spin-0 mass squared, while the lower
    panel shows the spin-$\frac12$ mass contribution, both in units of $q^2 m$.
    The dashed lines indicate the regions within which the finite-volume
    contributions are below $1\%$ relative to the electromagnetic corrections to
    the mass.}
    \label{fig:mfv}
\end{figure}

The $\bigo(\frac{1}{L})$-improved finite-volume corrections to the mass are
presented in~\cref{fig:mfv}. The improved effects are smaller for large values
of $mL$, but are, in fact, larger than in $\qedl$ for values around $mL=4$,
making the benefit of this improvement strategy limited. This provides an
additional motivation to look for better improvement prescriptions.
\begin{table}[t]
    \centering
    \begin{tabular}{|l|l|l|l|l|l|}
    \hline
    Improvement                                            & $w_1$        &
    $w_2$         & $c_2^{(w)}$   & $c_1^{(w)}$   & $c_0^{(w)}$  \\ \hline
    None                                                   & $0$          & $0$
    & $-8.91363292$ & $-2.83729748$ & $-1$         \\
    $\bigo(\frac{1}{L})$                                   & $1.48560549$ & $0$
    & $0$           & $6.07633544$  & $7.91363292$ \\
    cumulative $\bigo(\frac{1}{L^2})$                      & $0.86681632$ & $0$
    & $-3.71273496$ & $2.36360048$  & $4.20089796$ \\
    cumulative $\bigo(\frac{1}{L^3})$                      & $0.52392582$ & $0$
    & $-5.77007797$ & $0.30625747$  & $2.14355495$ \\
    cumulative $\bigo(\frac{1}{L^2})+\bigo(\frac{1}{L^3})$ & $2.04145881$ &
    $-0.93739607$ & $-2.28925650$ & $1.45738594$  & $0$          \\ \hline
    \end{tabular}
    \caption{Summary of improvement weight factors and finite-volume
    coefficients according to the improvement prescriptions for the mass of
    charged hadrons described in~\cref{sec:seirimp}. Values from cumulative
    improvement prescriptions are given for the reference scale $\mu_0=mL_0=4$.}
    \label{tab:ircoef}
\end{table}
\subsubsection{Cumulative improvement
\label{subsec:cumulative}}
From what was derived so far, it is reasonable to assume, even without prior
knowledge of the finite-volume coefficients, that $\Delta\omega_0(\psp)^2/m^2$
has a power expansion in
\begin{equation}
    \frac{c_{3-j}^{(w)}}{(2\pi)^{3-j}\mu^j}
    \label{eq:irimppower}
\end{equation}
with  ${\cal O}\left(1\right)$ coefficients. Consequently, one possible strategy
to circumvent the positivity issue encountered in the previous section is to
tune the improvement weights $w_n$ to obtain
\begin{equation}
    \frac{c_2^{(w)}}{4\pi^2\mu_0}+\frac{c_1^{(w)}}{2\pi\mu_0^2}=0\,,
    \label{eq:1p2irimpeq}
\end{equation}
for a reference volume $\mu_0=mL_0$. This is achievable by a minimal choice
\begin{equation}
    w_1=-\frac{\pi}{6}\frac{2+\mu_0}{2\pi+\mu_0}c_1\,.
    \label{eq:1p2irimp}
\end{equation}
In lattice QCD+QED calculations, typical values for $\mu$ are: $\mu \gtrsim 4$.
For any positive $\mu_0$,~\cref{eq:1p2irimp} gives $w_1>-1$, which does not
violate the positivity of the action. Numerical values for $w_1$ and the
$c_j^{(w)}$ coefficients are given in~\cref{tab:ircoef}. Similarly, combinations
of  the three first orders can be suppressed by solving
\begin{equation}
    \frac{c_2^{(w)}}{4\pi^2\mu_0}+\frac{c_1^{(w)}}{2\pi\mu_0^2}
    +\frac{c_0^{(w)}}{\mu_0^3}=0\,,
    \label{eq:1p2p3irimpeq}
\end{equation}
which gives
\begin{equation}
    w_1=\frac{\pi}{6}\frac{4\pi-2\mu_0c_1-\mu_0^2c_1}
    {4\pi^2+2\pi\mu_0+\mu_0^2}\,.
\end{equation}
Finally,~\cref{eq:1p2irimpeq,eq:1p2p3irimpeq} can be simultaneously solved
using the two weights $w_1$ and $w_2$ to obtain
\begin{align}
    w_1&=\frac{1}{6}\frac{2\sqrt{2}\pi+\mu_0+4\pi c_1+2\pi\mu_0c_1}
    {2\sqrt{2}\pi-4\pi-\mu_0},\\
    w_2&=-\frac{1}{6}\frac{2\pi+\mu_0+2\pi c_1+\pi\mu_0 c_1}
    {2\sqrt{2}\pi-4\pi-\mu_0}\,.
\end{align}
The finite-volume effects in the cumulative improvement at the reference scale
$\mu_0=4$ are shown in~\cref{fig:mfv}. The
$\bigo(\frac{1}{L^2})+\bigo(\frac{1}{L^3})$ cumulative improvement is
efficient, producing subpercent relative finite-volume corrections for any
$mL>4$ in the masses of both spin-$0$ and spin-$\frac12$ charged particles.
\subsubsection{Universality of the procedure}
One legitimate worry about the improvement procedure is its observable
dependence. The quality of the improvement is, in principle, determined by the
target observable and could actually enhance finite-volume effects on other
observables. However, the cumulative improvement scheme is based on minimizing
the first terms in the volume expansion assuming some naturalness of the power
expansion driven by~\cref{eq:irimppower}. Further investigations are needed to
determine the extent to which this is a good assumption. This likely
requires generalizing the formal derivation in~\cref{sec:masterfv} to arbitrary
one-loop diagram for leading-order electromagnetic corrections. Such a
calculation is beyond the scope of the present paper. Nevertheless, one of the
interesting aspects of the procedure presented here is to emphasis the
arbitrariness in the choice of scheme for subtracting the photon zero-mode. As
explained in~\cref{sec:qedl}, the standard $\qedl$ scheme is minimal with
respect to locality in time, but there is no reason for it to be optimal for
finite-volume effects. The improvement procedure discussed here gives a
practical example of the potential benefits of modifying this prescription.
\label{sec:latimp}

%% file: sections/lattice.tex
As a laboratory to test ideas presented in the previous sections and to allow
for checks of the finite-volume relations, we have performed a dedicated
numerical lattice QED study to compute the self-energy of a fundamental charged
scalar particle in a finite volume with periodic boundary conditions.
\subsection{Lattice scalar $\qedl$}
Consider a finite 4-dimensional lattice, $\Lambda^4$, with a lattice
spacing $a$, temporal extent $T=aN_T$ and spatial extent $L=aN_L$.
It is convenient to define a translation operator in the $\mu$ direction 
\begin{equation}
    \tau_{\mu}f(x)=f(x+a\hat{\mu})\,,
\end{equation}
where $f$ is an arbitrary function of coordinates and $\hat{\mu}$ is the unit
vector in direction $\mu$, using which  discrete derivatives and covariant
derivatives can be defined, 
\begin{align}
    \df_{\mu}&=a^{-1}(\tau_{\mu}-1),\\
    \db_{\mu}&=a^{-1}(1-\tau_{-\mu}),\\
    \Dt&={\textstyle\sum_{\mu}}\df_{\mu}^{}\db_{\mu}\,\\
    \cdf_{\mu}&=a^{-1}(e^{iaqA_{\mu}}\tau_{\mu}-1)=
    e^{iaqA_{\mu}}\df_{\mu}+a^{-1}(e^{iaqA_{\mu}}-1)\,,\\
    \cdb_{\mu}&=a^{-1}(1-\tau_{-\mu}e^{-iaqA_{\mu}})
    =\db_{\mu}e^{-iaqA_{\mu}}+a^{-1}(1-e^{-iaqA_{\mu}})\,,\\
    \cDt&={\textstyle\sum_{\mu}}\cdf_{\mu}\cdb_{\mu}\,,
\end{align}
where $A_{\mu}$ is the $\U(1)$ gauge potential and $q$ is the charge of the scalar particle.
\subsubsection{Lattice action and observables}
On such a lattice, the QED action for a scalar complex field in Feynman gauge is
given by
\begin{equation}
  S[\phi,A_{\mu}]=S_{\phi}[\phi,A_{\mu}]+S_{\mathrm{Feyn.}}[A_{\mu}]\,,
  \label{eq:latsqedact}
\end{equation}
where the matter term is
\begin{equation}
    S_{\phi}[\phi,A_{\mu}]=a^4\suml{x}\left\{{\textstyle\sum_{\mu}}
    |\cdf_{\mu}\phi(x)|^2+m^2|\phi(x)|^2\right\}
    =
    a^4\suml{x}\phi(x)\Delta\phi(x)^*
    \,,\label{eq:latphiact}
\end{equation}
with $\Delta=-\cDt+m^2$. The gauge action takes the form
\begin{equation}
  S_{\mathrm{Feyn.}}[A_{\mu}]=
  a^4\suml{x}\left\{\frac14{\textstyle\sum_{\mu,\nu}}F_{\mu\nu}(x)^2
    +\frac12{\textstyle\sum_{\mu}}[\df_{\mu}A_{\mu}(x)]^2\right\}
  =-\frac{a^4}{2}\suml{x}A_{\mu}(x)\Dt A_{\mu}(x)\,,
\end{equation}
where $F_{\mu\nu}=\df_{\mu}A_{\nu}-\df_{\nu}A_{\mu}$. 

In this theory, a scalar observable $O[\phi,\phi^*]$ has the expectation value
\begin{equation}
    \braket{O}=\frac{1}{\mathscr{Z}_{\mathrm{L}}}\int\Diff A_{\mu}\,
    \Diff\phi\,\Diff\phi^*\,
    O[\phi,\phi^*]\exp(-S_L[\phi,A_{\mu}])\,,
\end{equation}
where $\mathscr{Z}_{\mathrm{L}}$ is the partition function, and
the index $L$ indicates use of the the $\qedl$ prescription described
in~\cref{sec:qedl}, corresponding to the condition
\begin{equation}
    a^3\sum_{\mathbf{x}\in\Lambda^3}A_{\mu}(t,\mathbf{x})=0\,,
\end{equation}
where $\Lambda^3$ is the spatial sub-lattice. Since the action in
~\cref{eq:latphiact} is quadratic in $(\phi,\phi^*)$, the  integration over
the scalar fields can be performed analytically,
\begin{equation}
  \braket{O}=\frac{1}{\mathscr{Z}_{\mathrm{L}}}\int\Diff A_{\mu}\,
  O_{\mathrm{Wick}}[\Delta^{-1}]\det(\Delta)^{-\frac12}
  \exp(-S_{\mathrm{Feyn.,L}}[A_{\mu}])\,,\label{eq:wickpint}
\end{equation}
where $O_{\mathrm{Wick}}$ is the function arising from Wick contractions of
matter fields in operator $O$. Due to the symmetry $A_{\mu}\mapsto-A_{\mu}$ of
the action $S_{\mathrm{Feyn.,L}}$, any contributions that are odd in the charge
$q$ are absent from expectation values. To obtain the leading order,
$\bigo(q^2)$, corrections to $\braket{O}$, it is therefore sufficient to use the
quenched theory, \ie to set $\det(\Delta)=1$.~\cref{eq:wickpint} can then be
evaluated using Monte-Carlo techniques by computing the observable
$O_{\mathrm{Wick}}[\Delta^{-1}]$ for  $A_{\mu}$ fields sampled from  the
Gaussian distribution $\Diff A_{\mu}\,\exp(-S_{\mathrm{Feyn.,L}}[A_{\mu}])$.
\subsubsection{Scalar propagator}
The terms in the expansion of the lattice Laplacian $\cDt$ in leading powers of $q$ are
\begin{equation}
  \Delta=\Delta_0+q\Delta_1+q^2\Delta_2+\bigo(q^3)\,,
\end{equation}
with
\begin{align}
  \Delta_0&=-a^{-2}(\tau_{\mu}+\tau_{-\mu}-2)\,,\\
  \Delta_1&=-ia^{-1}
  \,{\textstyle\sum_{\mu}}(A_{\mu}\tau_{\mu}-\tau_{-\mu}A_{\mu})\,,
  \label{eq:d1def}\\
  \Delta_2&=\frac12
  \,{\textstyle\sum_{\mu}}(A_{\mu}^2\tau_{\mu}+\tau_{-\mu}A_{\mu}^2)\,.
  \label{eq:d2def}
\end{align}
The scalar propagator is then given by
\begin{equation}
  \Delta^{-1}=\Delta_0^{-1}-q\Delta_0^{-1}\Delta_1\Delta_0^{-1}
  +q^2\Delta_0^{-1}\Delta_1\Delta_0^{-1}\Delta_1\Delta_0^{-1}
    -q^2\Delta_0^{-1}\Delta_2\Delta_0^{-1}+\bigo(q^3)\, ,
\end{equation}
which can be diagramatically represented as 
\begin{equation}
  \raisebox{-8pt}{\includegraphics{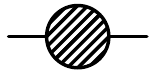}}
  =\raisebox{1pt}{\includegraphics{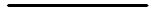}}-q\,
  \raisebox{-1pt}{\includegraphics{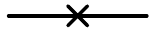}}
  +q^2\,\raisebox{-1pt}{\includegraphics{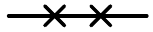}}
  -q^2\,\raisebox{-2pt}{\includegraphics{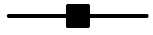}}\,.
  \label{eq:scalpropexp}
\end{equation}
Here, the line, cross and square vertices represent the free scalar propagator,
an insertion of $\Delta_1$ and an insertion of $\Delta_2$, respectively.

One may define the lattice Fourier transform,
\begin{equation}
[\mathcal{F}f(x)](k)= \tilde{f}(k) = a^4\suml{x} f(x) e^{-i k \cdotp x}
\end{equation}
and its inverse,
\begin{equation}
[\mathcal{F}^{-1}\tilde{f}(k)](x)=f(x) = \frac{1}{TL^3} \sum_{k \in\hat{\Lambda}^4} \tilde{f}(k) e^{i k \cdotp x}
\end{equation}
to represent the scalar propagator in momentum space. The free propagator
is given by
\begin{equation}
  \Delta_0^{-1}=\F^{-1}G\F
\end{equation}
where $G$ is the diagonal, momentum space
operator
\begin{equation}
  G(p)=\frac{1}{\hat{p}^2+m^2}
\end{equation}
with the lattice momentum $\hat{p}_{\mu}=\frac{2}{a}\sin\left(\frac{ap_{\mu}}{2}\right)$.
Further, using~\cref{eq:d1def,eq:d2def}, it is straightforward to show that in
momentum space
\begin{align}
  \raisebox{-1pt}{\includegraphics{feyn_prop_d1.pdf}}=
  \Delta_0^{-1}\Delta_1\Delta_0^{-1}&=-ia^{-1}\sum_{\mu}
  \F^{-1}(G\F A_{\mu}\F^{-1}\psi_{\mu}G-G\psi_{\mu}^*\F A_{\mu}\F^{-1}G)\F\,,\\
  \raisebox{-2pt}{\includegraphics{feyn_prop_d2.pdf}}=
  \Delta_0^{-1}\Delta_2\Delta_0^{-1}&=\frac12\sum_{\mu}
  \F^{-1}(G\F A_{\mu}^2\F^{-1}\psi_{\mu}G+G\psi_{\mu}^*\F A_{\mu}^2\F^{-1}G)\F
  \,,
\end{align}
where $\psi_{\mu}(p)=e^{iap_{\mu}}$. These expressions
are rather formal depictions of what could be obtained through the Feynman
rules of lattice scalar QED in a background electromagnetic field. In this form,
it is clear that the $\bigo(q^2)$ expansion of the inverse operator
$\Delta^{-1}$ in a given stochastic field $A_{\mu}(x)$ can be computed solely
using a fast Fourier transform (FFT) algorithm. This approach has two important
advantages compared to more conventional approaches that use iterative inverters, such as
the conjugate gradient algorithm. First, the complexity of performing a FFT is
independent of the  mass of the particle, and second, it scales as $V\log(V)$
where $V=TL^3$, making it of practical use for large-volume studies.
\subsection{On-shell self-energy from Euclidean-time correlators}
\label{sec:timemom}
The primary output of lattice QED simulations performed are time correlators,
from which we will obtain the particle's self-energy. The on-shell point is
defined through an analytic continuation of $\Sigma(p)$ to imaginary $p_0$. As
is usual in Euclidean field theory, this point is accessed through the
large-time behavior of relevant correlation functions. The 2-point function of
the charged scalar particle in the time-momentum representation is given by
\begin{equation}
  C(t,\psp)=a^3\sum_{\mathbf{x}\in\Lambda^3}
  \braket{\mathrm{T}[\phi(t,\mathbf{x})\phi(0)^{\dagger}]}
  e^{-i\psp\cdotp\mathbf{x}}\,.
  \label{eq:lat2pt}
\end{equation}
At small electric charges, this function can be decomposed into the tree level
and first-order electromagnetic corrections
\begin{equation}
  C(t,\psp)=C_0(t,\psp)+C_1(t,\psp)\,,
\end{equation}
which, in practice, are obtained from the diagrams in~\cref{eq:scalpropexp}. The
full spectral representation of $C_1(t,\psp)$ on a continuous space-time is
presented here, while the lattice equivalent of this result, which is used to
analyze computed lattice correlators, is obtained in~\cref{sec:latticetmom}.

The function $C_0(t,\psp)$ is the free scalar propagator, and is given by
\begin{equation}
  C_0(t,\psp)=\int
  \frac{\diff p_0}{2\pi}\frac{e^{ip_0t}}{p^2+m^2}
  =\frac{e^{-\om|t|}}{2\om}\,.
\end{equation}
The self-energy function $\Sigma(p)$ is defined by the
amputated first order corrections
\begin{equation}
  C_1(t,\psp)=\int\frac{\diff p_0}{2\pi}
  \frac{\Sigma(p)}{(p^2+m^2)^2}e^{ip_0t}\,,
  \label{eq:c1mom}
\end{equation}
as discussed in~\cref{eq:sedef}. It is defined through~\cref{eq:SigmaL,eq:K0}, and is given by 
\begin{equation}
  \Sigma(p)=\frac{q^2}{L^3}\sump_{\ksp\in\mathrm{BZ}(L)}
  \int\frac{\diff k_0}{2\pi}\left\{\frac{4}{k^2}
  -\frac{(2p-k)^2}{k^2[(p-k)^2+m^2]}\right\}\,,
\end{equation}
The $k_0$ integral can be performed to give
\begin{align}
 \Sigma(p)=\frac{q^2}{L^3}\sump_{\ksp\in\mathrm{BZ}(L)}
 &\left\{\frac{2}{|\ksp|}\right.+
 \frac{4p_0^2+\ksp^2+(2\psp-\ksp)^2}{2|\ksp|[p_0^2+\omega_{\gamma}(\psp,\ksp)^2]}
 \notag\\
 &\quad+\left.
 \frac{p_0^2+\omk^2+(2\psp-\ksp)^2}
 {2\omk[p_0^2+\omega_{\gamma}(\psp,\ksp)^2]}\right\}\,,
 \label{eq:contk0int}
\end{align}
which has the expected poles at $p_0=\pm i\omega_{\gamma}(\psp,\ksp)$, where
$\omega_{\gamma}(\psp,\ksp)=|\ksp|+\omk$ is the energy of a free photon-scalar
pair. Denoting contributions to~\cref{eq:c1mom} from the $p_0=i\om$ pole as
$C_{1,\Sigma}$, and those from $p_0=i\omega_{\gamma}(\psp,\ksp)$ poles as
$C_{1,\gamma}$, $C_1(t,\psp)$ can be split to
\begin{equation}
  C_1(t,\psp)=C_{1,\Sigma}(t,\psp)+C_{1,\gamma}(t,\psp)\,,
\end{equation}
where it can be shown that
\begin{align}
  C_{1,\Sigma}(t,\psp)&=\frac{e^{-\om|t|}}{4\om^3}\left\{[1+|t|\om]
  \Sigma(p_{\mathrm{o.s.}})
  -i\om\left . \frac{\partial\Sigma(p)}{\partial p_0}\right|_{p_{\mathrm{o.s.}}}\right\}\,,
  \label{eq:c1sig}\\
  C_{1,\gamma}(t,\psp)&= \frac{q^2}{L^3}\sump_{\ksp\in\mathrm{BZ}(L)}
  A(\psp,\ksp) e^{-\omega_{\gamma}(\psp,\ksp)|t|},
  \label{eq:c1ex}
\end{align}
where
\begin{equation}
 A(\psp,\ksp) \equiv -\frac{(2\psp-\ksp)^2-[2\omk+|\ksp|]^2}
  {4|\ksp|\omk[\omega_{\gamma}(\psp,\ksp)^2-\om^2]^2}\, .
  \end{equation}
Finally, an effective on-shell self-energy can be constructed from $C_0$ and $C_1$
correlators
\begin{align}
  \Sigma_{\mathrm{eff.}}(t)&\equiv 2q^2\om\frac{d}{d|t|}
  \left[\frac{C_1(t,\psp)}{C_0(t,\psp)}\right]
  \\
  &\hspace{-2.5ex}\underset{|t|\to+\infty}{=}\Sigma(p_{\mathrm{o.s.}})\,,
  \label{eq:effsig}
\end{align}
as previously obtained in Refs.~\citep{deDivitiis:2013xla,Boyle:2017gzv}. The
second term in the last line of~\cref{eq:effsig} represents contributions from a
tower of excited states, suppressed at large times by a decaying exponential of
the form $e^{-(\omega_{\gamma}(\psp,\ksp)-\om)|t|}$. The ground-state dominance
at large times relies entirely on the exponential suppression from the energy
gap $\omega_{\gamma}(\psp,\ksp)-\om$. This gap vanishes in the infinite volume
limit, creating the expected branch cut at the particle pole. This means that
large-volume lattice calculations of the effective self-energy are expected to
be severely contaminated by the excited spectrum. Further discussions of this
point will be presented in the next section by confronting the explicit
formula~\cref{eq:effsig} with the simulation data.
\subsection{Numerical results}
The strategy presented in the previous sections was implemented using the Grid
library~\citep{Boyle:2016lbp} to compute the time-momentum representation of the
charged scalar 2-point function.
\subsubsection{Simulation setup}
We calculated the 2-point function for a scalar field with bare mass $am=0.2$ on 
$12$ ensembles of $10000$ $\qedl$ gauge configurations with $12$ different
spatial volumes $12\le N_L\le 128$ and temporal extent $N_T=128$ or $N_T=256$,
and one ensemble of $3006$ $\qedl$ gauge configurations with volume
$192^3\times256$.
\subsubsection{Numerical extraction of the on-shell self-energy}
\begin{figure}[ht]
  \centering
  \includegraphics{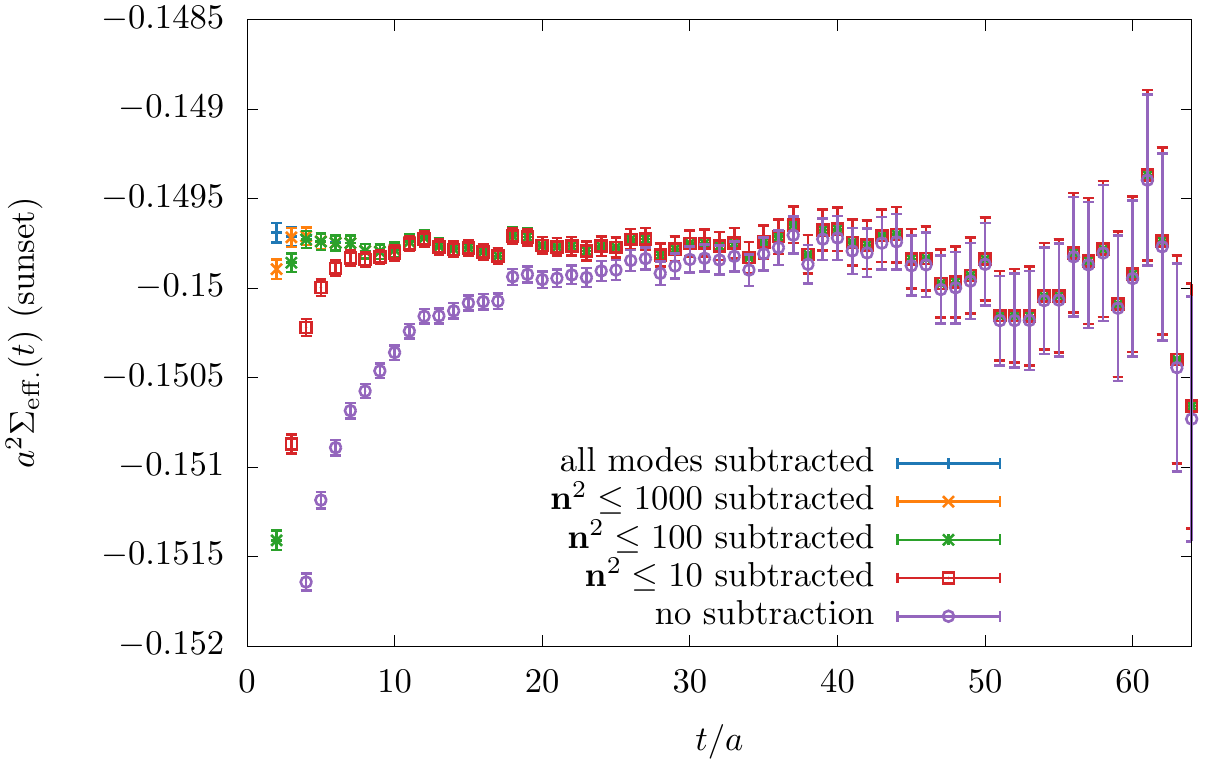}
  \caption{Sunset diagram (left diagram in~\cref{fig:feynman}) contribution to
  the effective scalar self-energy~\cref{eq:effsig} with various excited-state
  subtractions. The chosen volume here is $64^3\times 128$ and the momentum is
  $\psp=\frac{2\pi}{32a}(1,0,0)$. The subtractions are done using the
  spectral representation in~\cref{eq:effsiglattice} and a cutoff on the
  integer modes $\nsp=\frac{L}{2\pi}\ksp$.}
  \label{fig:sigmaeff}
\end{figure}

It was found to be essential to subtract excited-state contributions from the
$C_1(t,\psp)$ correlator in order to extract the on-shell self-energy from a fit
to the plateau region of the effective self-energy defined
in~\cref{eq:effsig}. For volumes $N_L\le64$, all $N_L^3-1$ excited states were
calculated analytically and subtracted. For larger volumes, to avoid calculation
of large numbers of excited states, excited states from all poles with
$|\ksp|^2\le\ksp_\mathrm{max}^2$ were subtracted. The threshold was chosen so
that halving $\ksp_\mathrm{max}^2$ would change
$\Sigma_\mathrm{eff.}(t_\mathrm{min})$ by less than one tenth of the statistical
uncertainty, where $t_\mathrm{min}$ is the lower limit of the fit interval.
\cref{tab:xstates} lists the number of excited states subtracted from each
scalar 2-point function. As an illustration, \cref{fig:sigmaeff} represents
results for the effective self-energy with various excited state subtractions.

After subtracting the excited-state contributions, the values of the on-shell
self-energy were extracted through a correlated fit to the plateau region of the
effective self-energy. Fit interval and number of excited states subtracted are
given for each volume and spatial momentum in~\cref{tab:fitranges}. In addition
to the statistical uncertainty from the ensemble average, the systematic
uncertainty arising from the choice of fit interval was estimated to be the
standard deviation of central values from fits to all sub-intervals with
$t_\mathrm{max}-t_\mathrm{min}\ge3$ and $p$-value $\ge0.05$.

It is important to notice that a full lattice QCD+QED calculation in a large
volume would suffer from the same significant contamination from excited states
with small energy gaps with the ground state. However, in such a setup, it is
not known how to extract the excited states that were obtained here
analytically. This suggests that, unless the way we extract energies from time
correlators is modified, performing QCD+QED simulations in large volumes will be
challenging.
\subsubsection{Numerical extraction of the off-shell self-energy}
For off-shell momenta accessible on the lattice, the scalar self-energy can
be calculated by dividing off the external free scalar propagators from the
first order corrections to the 2-point function:
\begin{equation}
\Sigma(p)=\left(\hat{p}^2+m^2\right)^2\sum_{t=0}^{T-1}C_1(t,\psp)e^{-ip_0t}.
\end{equation}
\label{sec:numosse}
\subsubsection{Signal-to-noise ratios in single-particle correlation functions}
The general results obtained by Parisi~\citep{Parisi:1983ae} and
Lepage~\citep{Lepage:1989hd} regarding the behavior of signal-to-noise (StN)
ratios in QCD correlation functions have been studied extensively, and are known
to correctly predict the behavior of the StN in (multi-)baryon correlation
functions, as detailed in \eg, Refs.~\citep{Beane:2009kya,Beane:2009gs}. They
are expected to apply with equivalent validity to the lattice QED correlation
functions given in~\cref{eq:lat2pt}. At late times, these correlation functions
are expected to behave as
\begin{equation}
C(t, {\bf p}) \rightarrow 
Z_{1} \ e^{-\omega({\bf p})  t}
\ +\ Z_{3} \
e^{- \omega_3({\bf p})  t}\ +\ \cdots \, ,
\end{equation}
where $\omega_3$ is the energy of three particles carrying a total momentum
${\bf p}$, the ellipses denote contributions from higher energy states,
including those with photons, and the $Z_i$ are overlap coefficients onto the
state $i$. The ``noise'' function is defined as the square root of the
(connected) variance correlation function which has late-time behavior
\begin{equation}
C_{\sigma^2}(t, {\bf p}) 
\rightarrow
Z_{\sigma^2; 0} \   e^{-2 \omega({\bf 0})  t}\ +\ 
Z_{\sigma^2; j} \  e^{-2 \omega^\prime_j ({\bf 0}) t}\ +\ \cdots \, ,
\label{eq:varcorB}
\end{equation}
where the interpolating operator has a non-zero overlap onto a pair of particles
at rest. Here, electromagnetic shifts in
the energies of multi-particle states have been neglected. The energy
$\omega^\prime_j $ appearing in~\cref{eq:varcorB} is that of the $j^{\rm th}$
excited state without disconnected contributions. In the absence of
interactions, the only state contributing to the noise correlation function is
one with back-to-back particles with momenta $\pm {\bf p}$. The StN ratios for
single-particle correlation functions are expected to degrade at late times as
\begin{equation}
C (t, {\bf p})/\sqrt{C_{\sigma^2}(t, {\bf p})}
\rightarrow
\tilde
Z_1 \ e^{-\omega_{StN}({\bf p}) t}
\ +\ \cdots 
\, ,
\end{equation}
where the StN energy scale appearing in the argument of the exponential is given by 
\begin{equation}
\omega_{StN}({\bf p})  
\ = \ 
\omega ({\bf p})  \ -\ \omega({\bf 0})  
\, ,
\label{eq:EStN}
\end{equation}
while being approximately independent of time at early times.
\begin{figure}[th]
  \centering
  \includegraphics{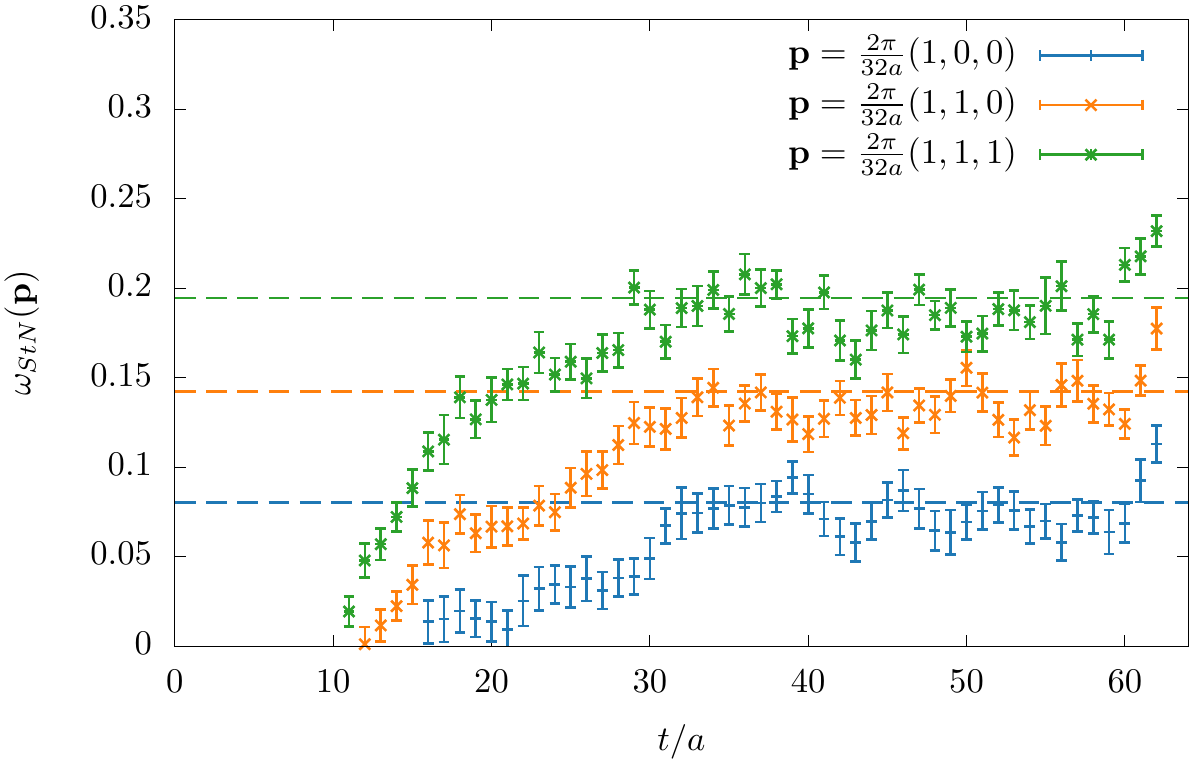}
  \caption{The energy scale, $\omega_{StN}({\bf p}) $, associated with the StN
  ratio, given in~\cref{eq:EStN}, calculated using the $32^3\times 128$
  ensemble. The points and uncertainties result from correlation functions of
  particles with boosts ${\bf p}=\frac{2\pi}{32a}\nsp$ with $\nsp = (1,0,0)$,
  $\nsp = (1,1,0)$, and $\nsp = (1,1,1)$. The dashed lines correspond to
  $\om-\omega(\mathbf{0})$ for these momenta.}
\label{fig:StN}
\end{figure}
The  results displayed in~\cref{fig:StN} show that the
Parisi-Lepage expressions (dashed horizontal lines) 
reproduce the late-time behavior of our results within uncertainties.

Generalizing to higher moments of the correlation functions, as has been done
previously for multi-baryon correlation functions~\citep{Beane:2014oea}, the
$n^{\rm th}$-even moments of the correlation functions can be argued to scale as
$\sim e^{-n \omega ({\bf 0}) t}$ at late times, while the $(n+1)^{\rm th}$-odd
moments scale as $\sim e^{- \omega_{n+1}({\bf p})t}$, where $\omega_{n+1}({\bf
p})$  is the minimum energy of $n+1$ $\phi$'s carrying momentum ${\bf p}$.
Consequently, at late times,  the boosted single-particle correlation functions
are expected to become symmetric and non-Gaussian.
\subsubsection{Finite-volume scaling}
In this section, the results of our scalar QED simulations are compared against
the analytical finite-volume effects determined in~\cref{sec:fv}. Specifically,
the scalar self-energy on the lattice was computed for several different spatial
volumes at fixed physical momenta. The infinite volume self-energy, calculated
in lattice perturbation theory, was subtracted from the lattice results and
compared with the analytical results given in~\cref{eq:os0,eq:offs0}.

\begin{figure}[t]
  \includegraphics{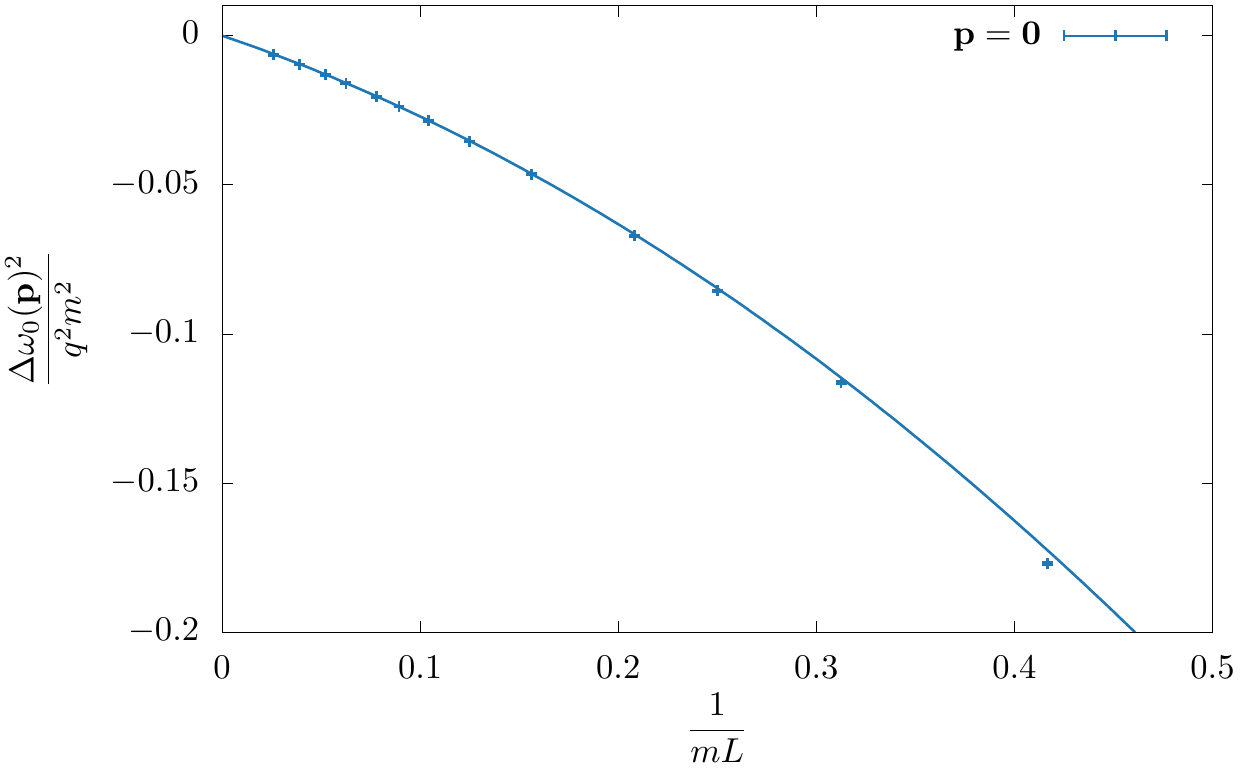}
  \caption{Volume scaling of the scalar on-shell self-energy in the rest frame.
  The points, and small associated uncertainties, come from the lattice scalar
  $\qedl$ simulations described in~\cref {sec:lattice} and the line corresponds to
  the analytical prediction~\cref{eq:os0}.}
  \label{fig:numfvos_rest}
\end{figure}

\begin{figure}[hp]
  \includegraphics{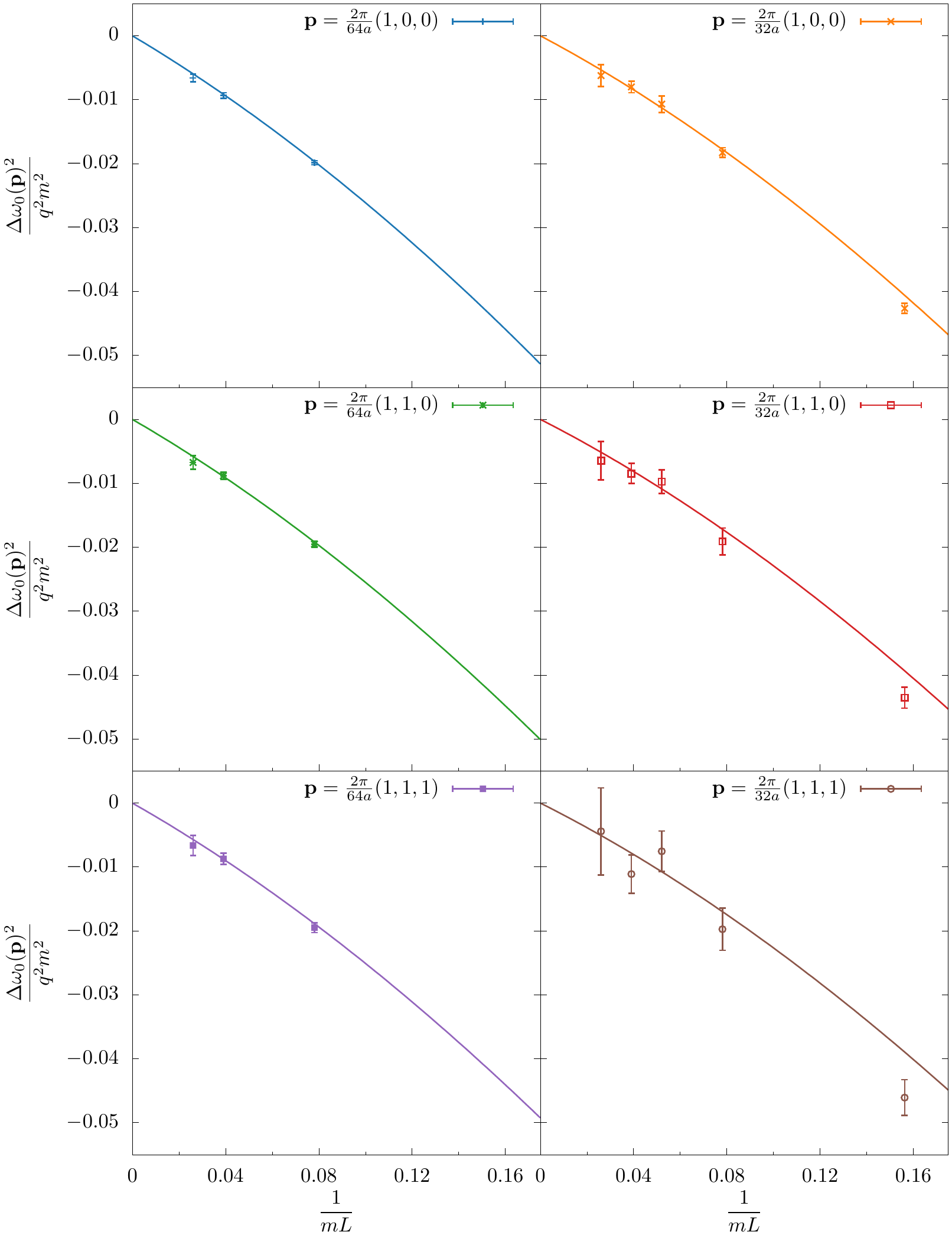}
  \caption{Volume scaling of the scalar on-shell self-energy for momenta of 
  various directions and magnitudes. Other details are identical 
  to~\cref{fig:numfvos_rest}.}
  \label{fig:numfvos_boosted}
\end{figure}

\begin{figure}[hp]
  \includegraphics{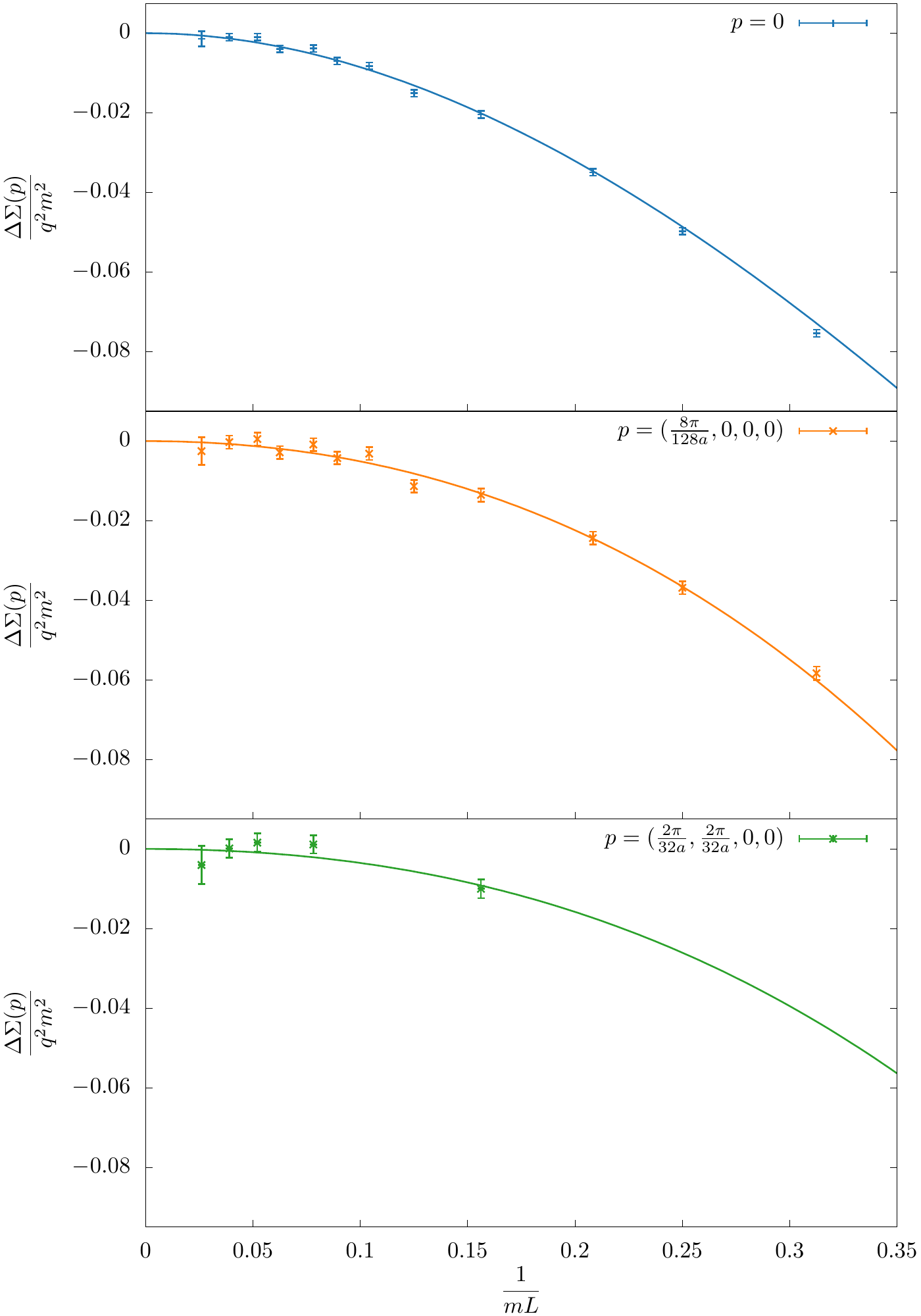}
  \caption{Volume scaling of the scalar self-energy for off-shell momenta with
  various orientations. The points come from the lattice scalar $\qedl$
  simulations described in~\cref {sec:lattice} and the line corresponds to the
  analytical prediction~\cref{eq:offs0}.}
  \label{fig:numfvoffs}
\end{figure}

For on-shell momenta, the volume scaling is shown for the rest frame
in~\cref{fig:numfvos_rest}, and for a selection of moving frames
in~\cref{fig:numfvos_boosted}. The lattice results are seen to agree with the
analytical results, except for small discrepancies at smaller volumes, which are
of $\bigo\left(e^{-mL}\right)$ and can therefore be attributed to exponential
effects neglected in the analytical calculation. By numerically reproducing representative data points in lattice perturbation theory in a finite volume, we have indeed confirmed that this discrepancy is related to the neglected
higher-order, exponentially suppressed finite-volume effects in our finite-volume expansion. For
$\psp=\frac{2\pi}{16a}(1,0,0)$ or larger, the poor StN ratio does not allow a
reliable extraction of the on-shell self-energy.  The volume scaling for a
selection of off-shell momenta is shown in~\cref{fig:numfvoffs}. Again, good
agreement is found between numerical and analytical calculations up to
exponential corrections.
\subsubsection{Infrared improvement}
The method of infrared improvement, described in~\cref{sec:irimp}, was
implemented in our numerical calculation. Improved gauge ensembles of 100
configurations were generated for each of the volumes, and for each choice of
the improvement weights given in~\cref{tab:ircoef}. The rest-frame scalar self
energy has been calculated on these improved ensembles, and checked through
exact analytical calculations of the difference in self-energy with and without
improvement.

\begin{figure}[t]
  \includegraphics{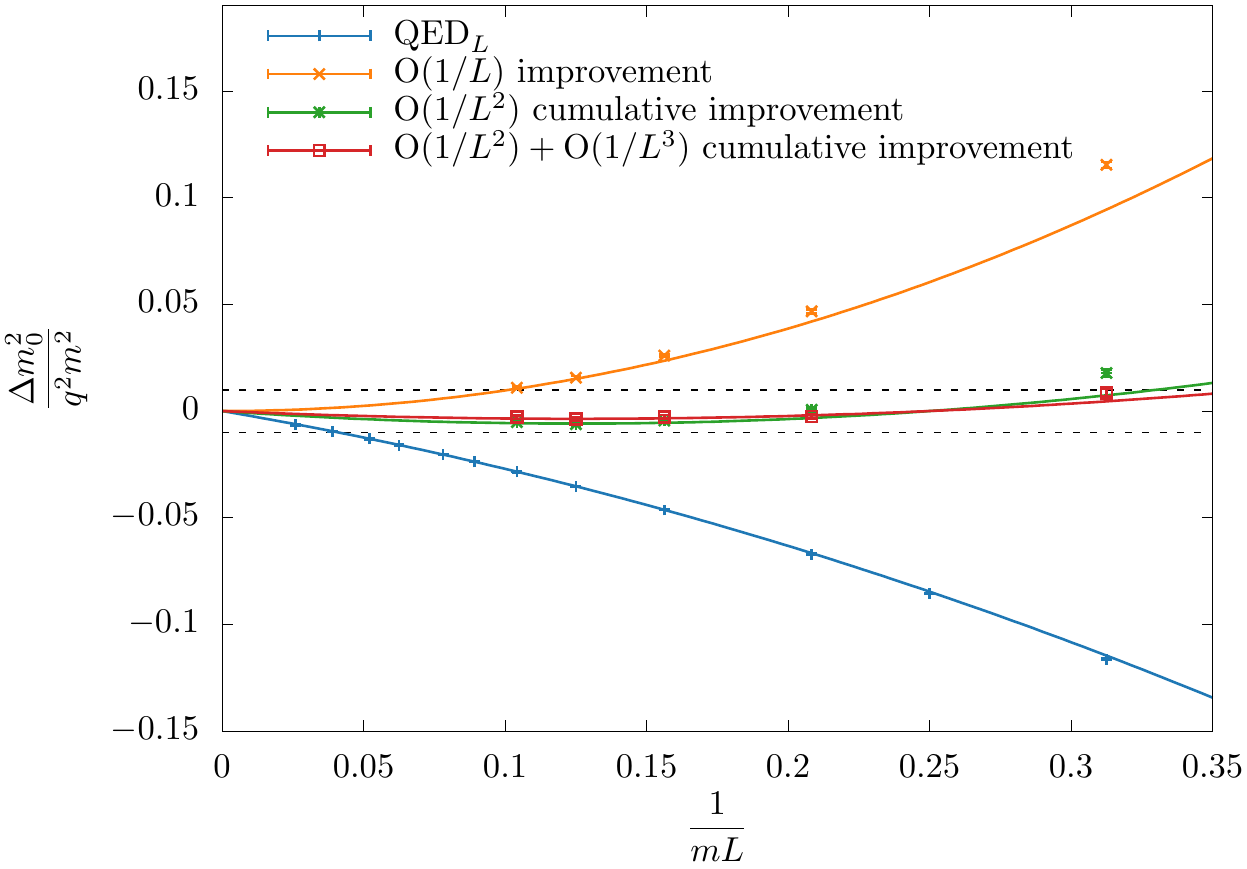}
  \caption{Relative finite-volume contributions to the mass squared of the
  scalar particle in $\qedl$, and the improved versions described
  in~\cref{sec:irimp}. The points correspond to lattice scalar $\QED$
  simulations described in~\cref {sec:lattice}, and with photon actions improved
  using the method described in~\cref{sec:irimp}. The lines are the analytical
  predictions shown in the upper panel of~\cref{fig:mfv}. The dashed line
  indicate the region within which the finite-volume contributions are below a
  percent of the electromagnetic corrections to the mass.}
  \label{fig:numirimp}
\end{figure}

The upper panel of~\cref{fig:mfv} is reproduced in~\cref{fig:numirimp},
including the numerical values of finite-volume corrections to the mass of the
scalar particle from the lattice simulations. The volume scaling from the
improved ensembles behaves according to the analytical predictions, up to small
deviations which can be attributed to exponential corrections that have been
neglected in the analytical calculation. The discrepancy between numerical and
analytical results is significantly smaller without improvement than with
improvement, which we checked explicitly for representative data points. It
appears that there is a suppression of exponential corrections that is broken by
the improvement procedure.

%% file: sections/eft.tex
The finite-volume modifications to the properties  of charged particles in a
lattice volume can be described by low-energy effective field theories.
Calculations of the finite-volume mass of fundamental and composite charged
scalars and fermions in non-relativistic QED
(NRQED)~\citep{Lepage:1987gg,Thacker:1990bm,Hill:2012rh} were performed by two
of the authors~\citep{Davoudi:2014qua}. Finite-volume corrections to the mass
calculated with NRQED were found to be in agreement with  those of QED for
scalar particles at leading order in $\alpha$, while  a discrepancy was found
between QED and NRQED  for fermions  at $\bigo(1/L^3)$~\citep{Fodor:2015pna}.
This discrepancy is disturbing and has generated a number of subsequent
investigations, \eg Ref.~\citep{Fodor:2015pna, Lee:2015rua}. In this section,
we show why the NRQED calculations of the $\alpha/L^3$ finite-volume
contribution to the mass of a charged fermion in Ref.~\citep{Davoudi:2014qua}
was incomplete, and explain why a residual mass term must be included in the
NRQED Lagrange density to recover the correct low-energy QED result. We also
extend these calculations to the self-energy of charged scalars and fermions
carrying  momentum. As the charged particles of interest can have arbitrary
momentum in the rest frame of the lattice, NRQED does not provide an appropriate
framework to calculate the low-energy properties of particles moving with a large
momentum and effective field theories similar to Heavy-Quark Effective Field
Theory
(HQET)~\citep{Politzer:1988bs,Georgi:1990um,Neubert:1993mb,Manohar:1997qy} and
Heavy-Baryon Chiral Perturbation Theory
(HB$\chi$PT)~\citep{Jenkins:1990jv,Jenkins:1991ne} are required.
\subsection{Heavy-Scalar QED }
Heavy-Scalar QED (HSQED) is the EFT describing the low-momentum interactions of
a charged scalar field, $\phi$, with the electromagnetic field after removing
the momentum associated with its classical trajectory. The HSQED Lagrange
density is in Minkowski space-time by 
\begin{equation}
{\cal L}[\phi]=\phi_u^\dagger iu\cdot D \phi_u-
{1\over 2 m_\phi}\phi_u^\dagger D_\perp^2\phi_u-
\delta m_{\phi,u}\  \phi_u^\dagger\phi_u
\,,
\end{equation}
where $D_\mu = \partial_\mu + iqA_\mu$, and where the field has been redefined
into the non-relativistic convention $\phi_u\rightarrow \phi_u/\sqrt{2 m_\phi}$,
and $\delta m_{\phi,u}$ is a residual mass. The full four-momentum of $\phi$ is
$p=m_\phi u+k$, where $u^2=u_0^2-|{\bf u}|^2=1$, and the phase associated with
the classical trajectory of $\phi$ in infinite volume has been removed, $\phi(x)
= e^{-i m_\phi u\cdot x} \phi_u(x)$, leaving a residual momentum $k$.
$D_\perp^\mu = D^\mu - u^\mu u\cdot D$ where the equations of
motion~\citep{Politzer:1980me} have been used. The components of the
four-velocity are related to ${\bf v}$ by $u=\gamma(|\vel|)(1,\vel)$. The
dynamics of the electromagnetic field, $A_\mu(x)$,  with the spatial zero mode
removed are detailed in~\cref{sec:qedl}. The appearance of a residual mass term,
$\delta m_{\phi,u} $, is at the heart of the present discussion and concerns the
discrepancy between previous calculations~\citep{Davoudi:2014qua,Fodor:2015pna}.
Removing the classical trajectory associated with the infinite-volume mass,
$m_\phi$, through the aforementioned phase redefinition, leads to a vanishing
residual mass in infinite volume. However, as we shall show through matching to
the result of the full theory (scalar QED), in finite volume this term is
non-vanishing at $\bigo(1/L^3)$ due to the removal of the spatial zero mode of
$A_\mu(x)$. Calculations of the finite-volume contributions to the on-shell self
energy up to NNLO in HSQED give
\begin{equation}
  \Delta\Sigma_0^{\rm HSEFT}(p)_{|\sigma=0}=
  q^2\left\{
  \om(1-|\vel|^2)\frac{c_{2,1}(\vel)}{4\pi^2 L}+\frac{c_1}{2\pi L^2}
  -\frac{|\vel|^2}{4\om L^3}\right\}+
  2 m_{\phi} \delta m_{\phi,u}
  \,.
  \label{eq:HSQED}
\end{equation}
The LO and NLO terms agree with the results in the full theory,  but the NNLO
loop contributions differ by  a factor of two. As a result, matching the full
and effective theories determines the residual mass to be 
\begin{equation}
  \delta m_{\phi,u}=
  -\frac{ q^2|\vel|^2}{8 m_{\phi}
  \om L^3}=-\frac{q^2|\mathbf{v}|^2}{8\gamma(|\mathbf{v}|)\mu^3}m_{\phi}\,.
\end{equation}
The residual mass vanishes in the rest frame,  in agreement with previous
calculations, but there is a  non-zero residual mass for moving charged scalars
that scales as $\sim 1/L^3$.
\subsection{Heavy-Fermion QED}
The construction of the Heavy-Fermion QED (HFQED) follows along the same lines
as for HSQED, but with the elimination of the lower components of the fermion
spinor leaving a two-component theory. The field redefinitions can be found in
previous literature, with the low-energy effective Lagrange density constructed
to high orders in both the $1/m_\psi$ and coupling expansion, see \eg
Refs.~\citep{Manohar:1997qy,Hill:2012rh}. The Lagrange density describing the
low-energy dynamics of the charged fermion is known to be
\begin{align}
{\cal L}[\overline{\psi},\psi] &=
\overline{\psi}_u\left[iu\cdot D-\delta m_{\psi,u}- 
{1\over 2 m_\psi} D_\perp^2-c_F\ {1\over 4 m_\psi}\sigma_{\alpha\beta}
F^{\alpha\beta}\right.\notag\\
&\qquad\quad\left.-c_D{1\over 8 m_\psi^2}
u^\alpha (D_\perp^\beta F_{\alpha\beta})+
i c_S\ {1\over 8 m_\psi^2}
u_\lambda\sigma_{\alpha\beta} \{D_\perp^\alpha , F^{\lambda\beta} \}
\right]
\psi_u\,,
 \end{align}
where the coefficients of the operators, obtained by matching to infinite-volume
QED, are $c_F=c_D=c_S=q$ at tree level, in which limit the residual mass
$\delta m_{\psi,u}$ vanishes.

Calculation of the finite-volume contribution to the 
fermion self-energy with HFQED gives
\begin{align}
  \Delta\Sigma_{\frac12}(p)_{|\sigma=0}&= 
  q^2\left\{\frac{m_{\psi} c_{2,1}(\vel)}{8\pi^2\om L}+
  \frac{c_1}{4\pi m_{\psi} L^2}+
  \frac{2\om^2+m_{\psi}^2}{8m_{\psi}\om^3L^3}
  \right\}
  \notag\\
  &\quad+
  \frac{1}{8m_{\psi}\om L^3} \left( 
  2 c_F^2+ q c_D-3q^2
  \right)+
  \delta m_{\psi,u}\,,
\end{align}
where the second to last  term vanishes with the tree-level matching conditions.
In order to recover the self-energy calculated in $\qedl$, given in~\cref{eq:oshalf},
\begin{equation}
\delta m_{\psi,u}=q^2\frac{2\om^2+m_{\psi}^2}{8m_{\psi}\om^3L^3}=
\frac{q^2(1+2\gamma(|\mathbf{v}|)^2)}{8\gamma(|\mathbf{v}|)^2\mu^3}m_{\psi}.
\end{equation}
The residual mass contribution adds to the loop
contribution in HFQED to recover the $\bigo(1/L^3)$ contribution calculated
with QED, by construction. Unlike the case of the charged scalar particle, the
residual mass associated with a charged fermion does not vanish in the rest
frame and its omission is seen to be responsible for the discrepancy in previous
calculations~\citep{Davoudi:2014qua,Fodor:2015pna}.

It has been previously argued  that contact interactions between fermions and
anti-fermions need to be included in the low-energy EFT in order to recover the
correct finite-volume QED mass shift at this order~\citep{Fodor:2015pna}. In the
rest frame, such interactions give rise to a contribution to the self-energy at
this order, enabling NRQED to reproduce QED without the need for a residual mass
term. One interpretation is that one must include anti-particles with a mass of
$\gamma= 2 m_{\psi}$ into the theory and contract the anti-particle operators to
recover this result. This is a somewhat unappealing feature of a low-energy EFT
as this introduces a mass scale of $2 m_{\psi}$ into the theory, and provides a
dynamical ultraviolet scale in loop integrals that obscure an order-by-order
power counting. The length scale of the anti-particle fluctuations is $1/(2
m_\psi)$, and through their interactions with the background charge density in
$J_\mu (x) - {1\over L^3}\int_{\T^3} d^3{\bf y}\ J_\mu (t,{\bf y})$, give rise
to a self-energy contribution of the form $1/(m_\psi^2 L^3)$. This makes clear
that the separation between ultraviolet and infrared lengths scales, that is
explicit in the construction of low-energy EFTs (particularly in matching to the
full theory), is explicitly violated by removing a spatial mode of the
electromagnetic field. In particular, the infinite-volume matching conditions
between QED and the low-energy EFTs should be modified by contributions of the
form $1/(m_\psi^3 L^3)$, which is found to be the case. As such contributions
arise from physics at the length scale set by $1/(2 m_\psi)$, they can be
included in the EFT through local counterterms as long as such length scales
are not probed in the EFT. The residual mass term in HFQED at this order in the
$\alpha$ expansion eliminates the need for such interactions with anti-particles
or with the background charge density. The physics described here is essentially
the same as that presented in Ref.~\citep{Patella:2017fgk} in which the details
of operator matching in $\lambda\phi^4$ theory was considered when the zero mode
of the field was removed.

We argue that from the calculational standpoint, QED and Scalar-QED are easier
to work with than HFQED and HSQED for fundamental particles given the
non-trivial finite-volume matching conditions. We anticipate that QED will be
the most effective framework to go  to higher orders in the loop expansion and
in the $1/L^n$ expansion. The complexity associated with the non-locality of QED
in the absence of the electromagnetic spatial zero-mode, and its implications
for matching between QED and low-energy EFTs, while apparently tractable, adds
new features to the EFTs that seem to be overly cumbersome.
\subsection{Implications for Hadrons and Composite Systems}
In light of what was presented in this section, it is natural to contemplate the
implications for hadronic theories, particularly Chiral Perturbation Theory
($\chi$PT),  HB$\chi$PT and nuclear EFTs. In these theories, contributions to
observables that are non-analytic in the quark masses are uniquely recovered
from quantum loops, while analytic contributions are generated by loops and
local counterterms in the Lagrange density. In the finite-volume QED, the
numerical values of all of the local counterterms are expected to be modified by
contributions scaling as $1/( \Lambda_\chi^3 L^3)$ due to the interactions of
the quarks with the background charge densities associated with the other quarks
and themselves. This is the same underlying mechanism that generates a non-zero
residual mass term in HSQED and HFQED. We conclude that, while $\chi$PT and
other low-energy EFTs can be used to determine the leading finite-volume
electromagnetic contributions, and used to extrapolate them away, addressing
contributions that scale as $1/L^3$ or higher appears to be more challenging.

%% file: sections/sco.tex
High-precision studies of strongly interacting hadronic systems using the
numerical technique of lattice QCD require that QED is also included as a
dynamical quantum field theory. Such studies are critical to the success of
several experimental efforts in both high-energy physics and nuclear physics,
including programs to measure anomalous magnetic moment of the muon and
CP-violating observables in the decay of select hadrons; investigations that aim
to find new physics by revealing minuscule deviations from the Standard Model
predictions. Recognizing the need to include QED, numerical technologies and
theoretical frameworks have been developed in recent years with which to
facilitate lattice QCD+QED calculations. Unlike QCD, in which the strong
dynamics confine the color charges of quarks and gluons, leading to a mass gap
in the spectrum of the theory, QED contains massless photons coupled to a
conserved charge, which introduce additional complications into the
implementation and analysis of lattice QCD+QED calculations. The complications
are the consequence of restricting QED to a finite spatial volume, where the
need to impose boundary conditions on the fields ``collides'' with the classical
equations of motion, including Gauss's law and Ampere's law. Perhaps the
simplest technique to deal with this problem is to eliminate the zero spatial
momentum mode of the photon field, and numerically evaluate observables in the
remaining non-local QED-like field theory, called $\qedl$. The penalties
incurred for such a modification to QED include power-law volume corrections to
observables and the loss of the standard lore for constructing low-energy
effective field theories. The locality of theory is restored in the
infinite-volume limit. While other local formulations, including introducing a
small photon mass~\cite{Endres:2015gda, Bussone:2017xkb} or using other boundary
conditions~\cite{Polley:1993bn,Wiese:1991ku,Kronfeld:1992ae,Kronfeld:1990qu,Lucini:2015hfa,Hansen:2018zre},
exist to define QED in a finite volume, the success of $\qedl$ in recent
precision hadron spectroscopy studies, such as in Ref.~\cite{Borsanyi:2014jba},
appears promising, and motivated us to investigate further a number of key
theoretical and numerical aspects of such a scheme, to clarify its limitations,
and to introduce improvement schemes.

In particular, by focusing on the dynamics of a single fundamental charged
particle in lattice QED calculations, this work:
\begin{itemize}
\item Extends previous work to systems that are moving in the spatial volume. A
systematic approach is taken to obtain a power-series expansion that allows
power-law finite-volume QED corrections to the self-energy function to be
obtained at leading order in $\alpha$ and to all orders in $\frac{1}{L}$. This
approach provides a suitable framework for generalizing the formalism to
composite charged particles. Rotational symmetry breaking effects due to the
motion of a charged particle in a cubic volume are identified at leading orders
in the $\frac{1}{L}$ expansion and the associated three-dimensional integer sums
are evaluated via an efficient procedure. 
\item Introduces a mode-weighting technique that systematically improves the
infrared scaling of self-energy of both spin-0 and spin-$1/2$ particles,
reducing the size of finite-volume corrections to the mass of hadrons at typical
volumes used in current lattice QCD+QED calculations. The generality of the
procedure and its potential advantage in future calculations are discussed.
\item Verifies the theoretical results obtained for the case of a fundamental
scalar particle through a dedicated numerical study. The purpose for
high-precision in this study was to reveal any potential effect that may not
have been accounted for within the theoretical finite-volume framework, and to
understand their origin. For boosted systems, the density of finite-volume
states near the single-particle mass shell increases with velocity, as well as
with the lattice volume. In scalar QED, such excited-state contributions are
calculated analytically at leading order in $\alpha$ and removed from the
lattice correlation functions, such that an identification of the self-energy
from earlier Euclidean times is possible. The origin of observed signal-to-noise
in boosted correlators is discussed, and is found consistent with the discussion
by Parisi~\citep{Parisi:1983ae} and Lepage~\citep{Lepage:1989hd}.
\item Resolves a discrepancy in the literature concerning the $1/L^3$
finite-volume contributions calculated with NRQED and QED. It is shown how to
account for missing contributions in effective field theory through introducing
a local operator, a residual mass term, whose coefficient can only be fixed by
matching to the full theory, $\qedl$.
\end{itemize}

We anticipate that the ideas presented in this work, along with the detailed
theoretical and numerical explorations of $\qedl$, will be beneficial in the
development and analysis of future high-precision lattice QCD+QED calculations
of quantities of importance to experiment.

%% file: sections/masterfv.tex
This appendix provides the details of the calculation in~\cref{sec:strategy}.
We start by computing explicitly the residues $r_{\gamma}(\ksp,p)$ and
$r_{m}(\ksp,p)$ defined in~\cref{eq:genres},
\begin{align}
  r_{\gamma}(\ksp,p)&=
  \frac{f((i|\ksp|,\ksp),p)}{2|\ksp|[(p_0-i|\ksp|)^2+\omk^2]}\,,\label{eq:resr1}\\
  r_{m}(\ksp,p)&=
  \frac{f((p_0+i\omk,\ksp),p)}{2\omk\{[p_0+i\omk]^2+|\ksp|^2\}}\,.
\end{align}
For the photon-pole effect $\Delta_{\gamma}(p)$, the on-shell
and off-shell cases must be distinguished. Indeed, in the former the
denominator of~\cref{eq:resr1} has an extra singularity at $\ksp^2=0$
\begin{equation}
  (p_0-i|\ksp|)^2+\omk^2=2\om|\ksp|(-i\sqrt{\sigma-1}-\vdku)
  +\sigma\om^2\,,\label{eq:p0denom}
\end{equation}
which is $\bigo(|\ksp|)$ at $\sigma=0$. Therefore, with the on-shell momentum
$p=p_{\mathrm{o.s.}}=(i\om,\psp)$,
\begin{equation}
  r_{\gamma}(\ksp,p_{\mathrm{o.s.}})=
  \frac{f((i|\ksp|,\ksp),p_{\mathrm{o.s.}})}{4|\ksp|^2\om(1-\vdku)}\,.
\end{equation}
Power-law finite-volume effects can be generated by this expression in two ways:
firstly through the singularity in the denominator at $\ksp^2=0$, and secondly
because of the eventual lack of smoothness of the numerator through its
dependence to $|\ksp|$. The expansion of $f((i|\ksp|,\ksp),p)$ given
in~\cref{eq:fexp} leads to
\begin{equation}
  \Delta_{\gamma}(p_{\mathrm{o.s.}})=
  \frac{f_0(p_{\mathrm{o.s.}})c_{2,1}(\vel)}{16\pi^2\om L}
  +\sum_{j=1}^{+\infty}\frac{\xi_{2-j,1,j}(p_{\mathrm{o.s.}})}
  {2^{4-j}\pi^{2-j}\om L^{1+j}}
  +\cdots\,.
\end{equation}
In the off-shell case, the denominator in~\cref{eq:p0denom} is not simply
proportional to $|\ksp|$ but also has a constant term. Writing
the geometric expansion
\begin{equation}
  \frac{1}{2|\ksp|[(p_0-i|\ksp|)^2+\omk^2]}=
  \frac{1}{2\sigma\om^2|\ksp|}\sum_{j=0}^{+\infty}
  (-1)^j\left[\frac{2(-i\sqrt{\sigma-1}-\vdku)}{\sigma\om}\right]^j|\ksp|^{j}
  \,,
\end{equation}
and multiplying by~\cref{eq:fexp} leads to
\begin{equation}
  r_{\gamma}(\ksp,p_{\mathrm{o.s.}})=\frac{1}{2\sigma\om^2}
  \sum_{j=0}^{+\infty}\left\{\sum_{r=0}^{j}
  \left[\frac{2(i\sqrt{\sigma-1}+\vdku)
  }{\sigma\om}\right]^rf_{j-r}(\hat{\ksp},p)
  \right\}|\ksp|^{j-1}\,.
\end{equation}
This last expression is quite cumbersome, and at this stage it is more useful to
simplify it further on a case-to-case basis. The resulting leading finite-volume
effect is
\begin{equation}
  \Delta_{\gamma}(p)=
  \frac{f_{0}(p)c_1}{4\pi\sigma\om^2L^2}
  +\left[
  -\frac{i\sqrt{\sigma-1}\,f_{0}(p)}{\sigma^2\om^3}+
  \frac{\xi_{0,0,1}(p)}{2\sigma\om^2}
  \right]\frac{1}{L^3}+\bigo\left(\frac{1}{L^4}\right)\,.
  \label{eq:resdeltag}
\end{equation}
Turning to the charged particle function $\Delta_m(p)$, as functions of
$\ksp$, $\omk$ and $|\ksp|^2$ are analytic, and
$r_m(\ksp,p)$ does not have singularities in $\ksp$, including at the on-shell
point $p_0=i\om$, $r_m(\ksp,p)$ is an analytic function of $\ksp$ and
\begin{equation}
  \Delta_m(p)
  =-\frac{r_m(\mathbf{0},p)}{L^3}+\cdots=
  -\frac{f(((i+\sqrt{\sigma-1})\om,\mathbf{0}),p)}
  {2(i+\sqrt{\sigma-1})^2\om^3L^3}+\cdots\,,
  \label{eq:resdeltam}
\end{equation}
where ellipsis denote exponentially suppressed finite-volume effects. Thus, this
residue from the massive particle-pole only contributes a $\bigo(\frac{1}{L^3})$
finite-volume effect coming from the zero-mode subtraction. It is worth nothing
that there is an arbitrariness in our choice of the sign of $p_0$ at the
on-shell point in Euclidean spacetime. While contributions from the photon and
the particle pole in~\cref{eq:resdeltag,eq:resdeltam} are dependent upon this
choice, the final result for the on-shell self energy,~\cref{eq:os0,eq:oshalf},
is insensitive to such an arbitrariness.

%% file: sections/cj.tex
In this appendix, we discuss the numerical computation of the finite-volume
coefficients $c_{j,k}(\vel)$ defined by~\cref{eq:cjkdef}, that we recall here for
convenience:
\begin{equation}
    c_{j,k}(\vel)=\Delta_{\nsp}'\left[
    \frac{1}{|\nsp|^{j}(1-\vdnu)^{k}}
    \right]\,.
\end{equation}
It is clear that $c_{j,k}(\vel)$ is finite only if $j<3$ because of the IR
singularity at $|\nsp|=0$. Evaluating $c_{j,k}(\vel)$ numerically is a
non-trivial task because it relies on cancellations between a sum
and an integral which both diverge. One possible strategy, inspired by Refs.~\citep{NIJBOER1957309,Takahasi:1973jt}, is to introduce a damping function as
follows:
\begin{equation}
    c_{j,k}(\vel)=\Delta_{\nsp}'
    \left[\frac{f(\eta\nsp)}{|\nsp|^{j}(1-\vdnu)^{k}}\right]+
    \Delta_{\nsp}'
    \left[\frac{1-f(\eta\nsp)}{|\nsp|^{j}(1-\vdnu)^{k}}\right]\,,
    \label{eq:ewald}
\end{equation}
where the function $f_{\eta}$ has the following properties:
\begin{enumerate}
    \item[(F1)] $f(\nsp)$ decays faster than any power of $|\nsp|$ at
    infinity.
    \item[(F2)] $f(\eta\nsp)$ converges to $1$ for $\eta\to 0$.
    \item[(F3)] $|\nsp|^{-j}[1-f(\eta\nsp)]$ is an infinitely differentiable
    function on $\R^3$.
\end{enumerate}
The assumption~(F1) guarantees that the first term in~\cref{eq:ewald} can be
easily and cheaply evaluated numerically as the difference of rapidly converging
sum and integral. Assumptions~(F2) and~(F3) guarantee, up to a constant, the
second term in~\cref{eq:ewald} vanishes faster than any power of $\eta$ for
$\eta\to0$. In practice, one chooses a suitable function for $f$ and looks for a
window where $\eta$ is small enough such that the second term in~\cref{eq:ewald}
is negligible and does not have to be computed, while is still large enough to
allow for a fast convergence of the first term.

Strongly inspired by Refs.~\citep{Tan:2007bg,Takahasi:1973jt}, we choose the function
\begin{equation}
    f(\nsp)=1-\left(\tanh\{\sinh[|\nsp|
    (1-\vdnu)^{\frac{k}{j+2}}]\}\right)^{j+2}\,.
\end{equation}
for which it is straightforward to show that properties ~(F1) and~(F2) are
satisfied. In fact, the decay rate of this function is doubly exponential (\ie
exponential of an exponential). However, the norm $|\nsp|$ is not differentiable
on $\R^3$ so~(F3) needs to be discussed further. One observes that $f$ has been
crafted specifically so that $|\nsp|^{-j}(1-f(\eta\nsp))$ satisfies the
following properties around the origin: firstly, it is non-singular in $|\nsp|$
and secondly it is even in $|\nsp|$. This ensures that
$|\nsp|^{-j}(1-f(\eta\nsp))$ can be expanded in infinitely differentiable, even
powers of $|\nsp|$ for $|\nsp|\to 0$, and therefore~(F3) is true. More
explicitly, the following Taylor expansion around the origin can be derived,
\begin{equation}
    \frac{1-f(\eta\nsp)}{|\nsp|^{j}(1-\vdnu)^{k}}=\eta^{j+2}|\nsp|^2
    -\frac{j+2}{6}\eta^{j+4}(1-\vdnu)^{\frac{2k}{j+2}}|\nsp|^4
    +\bigo(|\nsp|^6)\,.
    \label{eq:cjkremexp}
\end{equation}

Sketching the evaluation of $c_{j,k}(\vel)$ using this specific damping
function, it is convenient to start by evaluating the the first term
of~\cref{eq:ewald} as the difference between a convergent sum and an integral,
\begin{equation}
    \Delta_{\nsp}'
    \left[\frac{f(\eta\nsp)}{|\nsp|^{j}(1-\vdnu)^{k}}\right]
    =\sump_{\nsp}\frac{f(\eta\nsp)}{|\nsp|^{j}(1-\vdnu)^{k}}
    -\int\diff^3\nsp\frac{f(\eta\nsp)}{|\nsp|^{j}(1-\vdnu)^{k}}\,.
\end{equation}
Because of its doubly-exponential rate of convergence, the sum is trivial to
evaluate numerically. The integral can be easily reduced to a one-dimensional
integral,
\begin{equation}
    \int\diff^3\nsp\,\frac{f(\eta\nsp)}{|\nsp|^{j}(1-\vdnu)^{k}}=
    4\pi\eta^{j-3}R_jA_{\frac{5k}{j+2}}(|\vel|)\,,
\end{equation}
with
\begin{equation}
    R_j=\int_0^{+\infty}\diff r\,\frac{1-\tanh[\sinh(r)]^{j+2}}{r^{j-2}}\,,
    \label{eq:rjdef}
\end{equation}
and the function $A_k$ is defined in~\cref{eq:Adef}.  Using the
expansion in~\cref{eq:cjkremexp}, it is clear that
\begin{equation}
    \Delta_{\nsp}'
    \left[\frac{1-f(\eta\nsp)}{|\nsp|^{j}(1-\vdnu)^{k}}\right]=0 +\cdots\,,
\end{equation}
where the ellipsis represents corrections that vanish exponentially fast as
$\eta\to 0$.

Finally, $c_{j,k}(\vel)$ can be written as
\begin{equation}
    c_{j,k}(\vel)=\sump_{\nsp}\frac{f(\eta\nsp)}{|\nsp|^{j}(1-\vdnu)^{k}}
    -4\pi\eta^{j-3}R_jA_{\frac{5k}{j+2}}(|\vel|)+\cdots\,,
\end{equation}
where only the sum and $R_j$ have to be evaluated numerically, both of which are
straightforward tasks given their doubly-exponential convergence rate.

%% file: sections/aklm.tex
In this appendix, we prove that the harmonic coefficient $a_{klm}(\vel)$
in~\cref{eq:ylmexp} is an $\mathcal{O}(|\vel|^l)$ quantity. These coefficients
are defined by
\begin{equation}
  a_{klm}(\vel)=\int_{\mathrm{S}^2}\diff^2\usp\,
  \frac{Y_{lm}(\theta_{\nsp},\phi_{\nsp})^*}{(1-\vdnu)^k}\,.
  \label{eq:aklmdef}
\end{equation}
The denominator of the integrand can be written as the power series
\begin{equation}
  \frac{1}{(1-\vdnu)^k}=\sum_{r=0}^{+\infty}
  \binom{k-1+r}{k-1}(\vudnu)^r|\vel|^r,
  \label{eq:geoexp}
\end{equation}
where the $\binom{n}{k}$ are the binomial coefficients. Further, $(\vudnu)^r$
can be written in the Legendre polynomial basis
\begin{equation}
  (\vudnu)^r=\sum_{s=0}^rp_{rs}P_s(\vudnu)\,,
\end{equation}
where it is known that~\citep{Weisstein} 
\begin{equation}
  p_{rs}=
  \begin{cases}
    \displaystyle
    \frac{(2s+1)r!}{2^{\frac{r-s}{2}} (\frac{r-s}{2})!(r+s+1)!!}& \text{if }
    r\geq s \text{ and } r\equiv s~(\mathrm{mod}~2),\\
    0              & \text{otherwise.}
  \end{cases}
\end{equation}
Using the spherical harmonics addition theorem
\begin{equation}
  P_s(\vudnu)=\frac{4\pi}{2s+1}\sum_{t=-s}^s
  Y_{st}(\theta_{\nsp},\phi_{\nsp})Y_{st}(\theta_{\vel},\phi_{\vel})^*\,,
\end{equation}
and the orthonormality of spherical harmonics, one obtains
\begin{equation}
  \int_{\mathrm{S}^2}\diff^2\usp\,
  (\vudnu)^rY_{lm}(\theta_{\nsp},\phi_{\nsp})^*=
  \begin{cases}
    \displaystyle
    \frac{4\pi }{2l+1}p_{rl}Y_{lm}(\theta_{\vel},\phi_{\vel})^* &\text{if }l\leq
    r\\
    0& \text{otherwise.}
  \end{cases}
\end{equation}
Finally, using this last result with~\cref{eq:aklmdef,eq:geoexp},  the
following power-series representation of $a_{klm}(\vel)$ is obtained
\begin{equation}
  a_{klm}(\vel)=
  \frac{4\pi }{2l+1} Y_{lm}(\theta_{\vel},\phi_{\vel})^* \sum_{r=l}^{+\infty}
  \binom{k-1+r}{k-1}p_{rl}|\vel|^r\,.
\end{equation}
This demonstrates that the rotational symmetry breaking effects with multipole
index $l$ are suppressed by a factor $|\vel|^l$, and explains why the
$c_{j,k}(\vel)$ coefficients are essentially equal to the symmetric result at low
velocities.

%% file: sections/latticetmom.tex
\begin{figure}[ht]
    \includegraphics{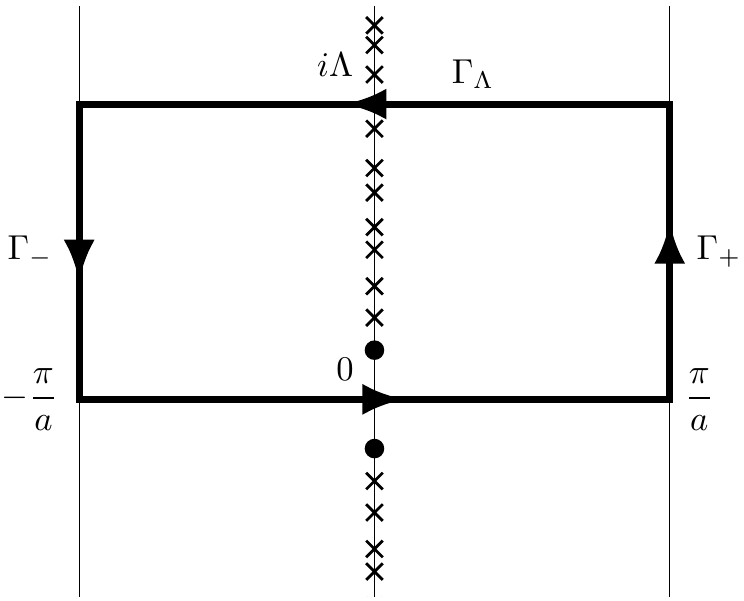}
    \caption{Rectangular contour used in~\cref{sec:latticetmom} for computing
    the energy Fourier transform of lattice correlators. This is a contour in
    the $p_0$ band with $-\frac{\pi}{a}\leq\Re(p_0)\leq\frac{\pi}{a}$. The
    integrals on the intervals $\Gamma_+$ and $\Gamma_-$ cancel by periodicity
    and the integral on $\Gamma_{\Lambda}$ decays exponentially fast when the
    cutoff $\Lambda$ goes to $+\infty$. The dots represent the poles contributing to the
    self-energy contribution~\cref{eq:C1SigmaLattice} and the crosses denote the
    scalar-photon scattering states contributing to~\cref{eq:C1gammalattice}.}
    \label{fig:latcont}
\end{figure}

The time-momentum representation of the self-energy function that was derived in
the continuum in~\cref{eq:c1sig,eq:c1ex} is extended to a self-energy function
defined on a cubic lattice. Here, the time extent is assumed to be infinite
while the spatial extent along each Cartesian coordinate is finite and has
length $L$. We start by the following definition of various lattice versions of
the energy and momentum
\begin{align}
    \hat{p}_{\mu} & =\frac{2}{a}\sin\left(\frac{ap_{\mu}}{2}\right)\,,\\
    \omega(\hat{\psp}) & =\sqrt{\hat{\psp}^{2}+m^{2}}\,,\\
    \omega_{\gamma}(\hat{\ksp}) & =|\hat{\ksp}|\,,\\
    \hat{\omega}(\psp) & =\frac{2}{a}\arcsinh
    \left[\frac{a\omega(\hat{\psp})}{2}\right]\,,\\
    \bar{\omega}(\psp) & =\frac{1}{a}\sinh
    \left[a\hat{\omega}(\psp)\right]
    =\omega(\hat{\psp})\sqrt{1+\left(\frac{a\omega(\hat{\psp})}{2}\right)^{2}}\,.
\end{align}
The scalar time-momentum correlator is defined as 
\begin{equation}
C\left(t,\mathbf{p}\right)=C_{0}\left(t,\mathbf{p}\right)+C_{1}\left(t,\mathbf{p}\right)\,,
\end{equation}
where $C_{0}\left(t,\mathbf{p}\right)$ is the free
lattice scalar correlator 
\begin{equation}
C_{0}\left(t,\mathbf{p}\right)=\int_{-\frac{\pi}{a}}^{\frac{\pi}{a}}\frac{dp_{0}}{2\pi}\frac{e^{ip_{0}t}}{\hat{p}_{0}^{2}+\omega(\hat{\psp})^{2}}=\frac{e^{-\hat{\omega}(\psp)|t|}}{2\bar{\omega}(\psp)}\,,\label{eq:C0}
\end{equation}
and the self-energy $\Sigma(p)$ is given through
the amputated first-order corrections 
\begin{equation}
C_{1}\left(t,\mathbf{p}\right)=\int_{-\frac{\pi}{a}}^{\frac{\pi}{a}}\frac{dp_{0}}{2\pi}\frac{\Sigma\left(p\right)}{\left(\hat{p}^2+m^2\right)^{2}}e^{ip_{0}t}\,,\label{eq:latticeC1}
\end{equation}
with
\begin{equation}
\Sigma\left(p\right)=\frac{q^{2}}{L^{3}}\sump_{\ksp\in\hat{\Lambda}^3}\int_{-\frac{\pi}{a}}^{\frac{\pi}{a}}\frac{dk_{0}}{2\pi}\left\{\frac{4-\frac{a^2}{2}\hat{p}^2}{\hat{k}^2}-\frac{(\widehat{2p-k})^2}{\hat{k}^2\left[(\widehat{p-k})^{2}+m^2\right]}\right\}\,.
\end{equation}
The $k_{0}$ integral can be performed to give 
\begin{align}
\Sigma(p)=
\frac{q^{2}}{L^{3}}\sump_{\ksp\in\hat{\Lambda}^3}
&\left\{\frac{2-\frac{a^2}{4}\hat{p}^2}{\bar{\omega}_{\gamma}(\ksp)}
-\frac{\frac{4}{a^2}\sin[ap_{0}-\frac{ia}{2}\hat{\omega}_{\gamma}(\ksp)]^{2}+(\widehat{2\psp-\ksp})^{2}}{2\bar{\omega}_{\gamma}(\ksp)[\frac{4}{a^2}\sin(\frac{a}{2}p_{0}-\frac{ia}{2}\hat{\omega}_{\gamma}(\ksp))^{2}+\omega(\widehat{\psp-\ksp})^{2}]}\right.\notag\\
&\quad\left.-\frac{\frac{4}{a^2}\sin[\frac{a}2p_{0}-\frac{ia}{2}\hat{\omega}(\psp-\ksp)]^{2}+(\widehat{2\psp-\ksp})^{2}}{2\bar{\omega}(\psp-\ksp)[\frac{4}{a^2}\sin(\frac{a}{2}p_{0}+\frac{ia}{2}\hat{\omega}(\psp-\ksp))^{2}+|\hat{\ksp}|^{2}]}\right\}\,.
\label{eq:After-k0-integral}
\end{align}
The terms in the last expression have poles in the upper plane at the
scalar-photon two-particle energy
$p_0=i\hat{\omega}_{\gamma}(\psp,\ksp)=i\hat{\omega}_{\gamma}(\ksp)+i\hat{\omega}(\psp-\ksp)$.
Note that analogous to the continuum case in~\cref{eq:After-k0-integral},
$p_0=\pm i[\hat{\omega}_{\gamma}(\ksp)-\hat{\omega}(\psp-\ksp)]$ is a removable
singularity of the function (\ie it is a pole with a vanishing residue). One
can now compute~\cref{eq:latticeC1} using the rectangular contour in $p_{0}$
described in~\cref{fig:latcont} to obtain 
\begin{equation}
C_{1}(t,\psp)=C_{1,\Sigma}(t,\psp)+C_{1,\gamma}(t,\psp)\,,
\end{equation}
where $C_{1,\Sigma}$ denotes the contribution from
the double pole at $p_{0}=i\hat{\omega}(\psp)$ and $C_{1,\gamma}$
denotes the contributions from the single pole $p_0=i\omega_{\gamma}(\psp,\ksp)$,
\begin{align}
    C_{1,\Sigma}(t,\psp) & =
    \frac{e^{-\hat{\omega}(\psp)|t|}}{4\bar{\omega}(\psp)^{3}}
    \left[\left(1+\frac{1}{2}a^{2}\omega(\hat{\psp})^{2}
    +\bar{\omega}(\psp)|t|\right)
    \Sigma\left(p_{\mathrm{o.s.}}\right)
    -i\bar{\omega}\left(\psp\right)
    \left.\frac{\partial\Sigma(p)}{\partial p_0}\right|_{p_{\mathrm{o.s.}}}
    \right]\,,\label{eq:C1SigmaLattice}\\
    C_{1,\gamma}(t,\psp) & =\frac{q^{2}}{L^{3}}\sump_{\ksp\in\mathrm{BZ}(L)}
    A(\psp,\mathbf{k})e^{-\hat{\omega}_{\gamma}(\psp,\ksp)|t|}\,,
    \label{eq:C1gammalattice}
\end{align}
where $p_{\mathrm{o.s.}}=(i\hat{\omega}(\psp),\psp$) is the on-shell momentum,
recovering $\hat{p}_{\mathrm{o.s.}}^{2}=-m^{2}$, and $A(\psp,\mathbf{k})$ is the
amplitude given by
\begin{equation}
    A(\psp,\mathbf{k})=
    -\frac{(\widehat{2\psp-\ksp})^{2}
    -\frac{4}{a^2}\sinh[a\hat{\omega}(\psp-\ksp)
    +\frac{a}{2}\hat{\omega}_{\gamma}(\ksp)]^{2}}
    {4\bar{\omega}(\psp-\ksp)\bar{\omega}_{\gamma}(\ksp)
    \{\frac{4}{a^2}\sinh[\frac{a}{2}\hat{\omega}_{\gamma}(\psp,\ksp)]^2
    -\omega(\hat{\psp})^{2}\}^{2}}\,.\label{eq:Alattice}
\end{equation}
Finally, an effective on-shell self-energy can be constructed from $C_0$ and $C_1$
correlators
\begin{align}
    \Sigma_{\mathrm{eff.}}(t,\psp) &=2q^{2}\bar{\omega}(\psp)\frac{\sign(t)}{a}
    \left[\frac{C_{1}\left(t+a,\psp\right)}{C_{0}\left(t+a,\psp\right)}
    -\frac{C_{1}\left(t,\psp\right)}{C_{0}\left(t,\psp\right)}\right]\\
    &\hspace{-2.4ex}\underset{|t|\to+\infty}{=}\Sigma(p_{\mathrm{o.s.}})\,.
    \label{eq:effsiglattice}
\end{align}
It is straightforward to verify that these results recover the continuum results
in~\cref{sec:lattice}.

%% file: sections/emfit.tex
The parameters used in extracting the scalar self-energy from the results of the
lattice calculations of this work, as described in~\cref{sec:numosse}, are
presented in~\cref{tab:xstates,tab:fitranges}.
\begin{table}[hp]
  \centering
  \begin{tabular}{|c|c|c|c|}
  \hline
  $N_L$ & $\frac{L}{2\pi}\psp$ & $\nsp_\mathrm{max}^2$ & $N_{\mathrm{sub.}}$ \\ \hline
  $\le 64$ &    & & $N_L^3-1$ \\
  80 & (0,0,0) & 128 &   6042  \\
  96 & (0,0,0) & 256 &  17076  \\
  96 & (3,0,0) & 256 &  17076  \\
  96 & (3,3,0) & 256 &  17076  \\
  96 & (3,3,3) & 512 &  48500  \\
  128 & (0,0,0) & 256 &  17076  \\
  128 & (2,0,0) & 512 &  48500  \\
  128 & (2,2,0) & 512 &  48500  \\
  128 & (2,2,2) & 512 &  48500  \\
  128 & (4,0,0) & 512 &  48500  \\
  128 & (4,4,0) & 512 &  48500  \\
  128 & (4,4,4) & 512 &  48500  \\
  192 & (0,0,0) & 512 &  48500  \\
  192 & (3,0,0) & 512 &  48500  \\
  192 & (3,3,0) & 1024 & 137064  \\
  192 & (3,3,3) & 1024 & 137064  \\
  192 & (6,0,0) & 1024 & 137064  \\
  192 & (6,6,0) & 1024 & 137064  \\
  192 & (6,6,6) & 1024 & 137064  \\ \hline
  \end{tabular}
  \caption{Number of excited states $N_{\mathrm{sub.}}$ subtracted from each
  scalar 2-point function. When applied, the cutoff imposed on the photon modes
  $\nsp_\mathrm{max}^2=\frac{L}{2\pi}\ksp_\mathrm{max}^2$ is given.}
  \label{tab:xstates}
\end{table}
\begin{table}[hp]
  \centering
  \begin{tabular}[t]{|c|c|c|c|c|c|c|}
  \hline
  $N_L$ & $\frac{L}{2\pi}\psp$  & $t_\mathrm{min}$ & $t_\mathrm{max}$ \\ \hline
  12 & (0,0,0) & 3 &  57 \\
  16 & (0,0,0) & 3 &  57 \\
  20 & (0,0,0) & 3 &  60 \\
  24 & (0,0,0) & 3 &  62 \\
  32 & (0,0,0) & 3 &  61 \\
  32 & (1,0,0) & 3 &  29 \\
  32 & (1,1,0) & 3 &  17 \\
  32 & (1,1,1) & 3 &  10 \\
  40 & (0,0,0) & 3 &  58 \\
  48 & (0,0,0) & 3 &  58 \\
  56 & (0,0,0) & 3 &  61 \\
  64 & (0,0,0) & 3 &  48 \\
  64 & (1,0,0) & 3 &  48 \\
  64 & (1,1,0) & 3 &  48 \\
  64 & (1,1,1) & 3 &  40 \\
  64 & (2,0,0) & 3 &  26 \\
  64 & (2,2,0) & 3 &  26 \\
  64 & (2,2,2) & 3 &  14 \\
  80 & (0,0,0) & 3 &  53 \\ \hline
  \end{tabular}
  \begin{tabular}[t]{|c|c|c|c|}
  \hline
  $N_L$ & $\frac{L}{2\pi}\psp$  & $t_\mathrm{min}$ & $t_\mathrm{max}$ \\ \hline
  96 & (0,0,0) & 3 &  59 \\
  96 & (3,0,0) & 3 &  34 \\
  96 & (3,3,0) & 3 &  19 \\
  96 & (3,3,3) & 3 &  9 \\
  128 & (0,0,0) & 3 & 107 \\
  128 & (2,0,0) & 3 & 102 \\
  128 & (2,2,0) & 3 &  29 \\
  128 & (2,2,2) & 3 &  30 \\
  128 & (4,0,0) & 3 &  31 \\
  128 & (4,4,0) & 3 &  19 \\
  128 & (4,4,4) & 3 &  13 \\
  192 & (0,0,0) & 3 & 110 \\
  192 & (3,0,0) & 3 &  94 \\
  192 & (3,3,0) & 3 &  57 \\
  192 & (3,3,3) & 3 &  31 \\
  192 & (6,0,0) & 3 &  11 \\
  192 & (6,6,0) & 3 &  11 \\
  192 & (6,6,6) & 3 &  10 \\ \hline
  \end{tabular}
  \caption{Time intervals used for fits to effective on-shell self energies.}
  \label{tab:fitranges}
\end{table}